\pdfoutput=1
\documentclass[10pt,journal,compsoc]{IEEEtran}

\ifCLASSOPTIONcompsoc
  \usepackage[nocompress]{cite}
\else
  \usepackage{cite}
\fi

\usepackage[hypertexnames=false]{hyperref}
\usepackage[T1]{fontenc}
\usepackage[utf8]{inputenc}
\usepackage[toc=false]{glossaries-extra}
\usepackage[spacing=true]{microtype}
\usepackage{graphicx}
\usepackage{xspace}
\usepackage{listings}
\usepackage{booktabs}
\usepackage{multirow}
\usepackage{pgf}
\usepackage{tikz}
\usepackage{tikz-uml}
\usepackage{caption}
\usepackage{subcaption}
\usepackage[binary-units]{siunitx}
\usepackage{multicol}
\usepackage{fnpct}
\usepackage{amsmath}
\usepackage{cleveref}
\usepackage[scaled=0.86]{beramono} 
\usepackage[inline]{enumitem}
\usepackage{amsthm}
\usepackage{colortbl}

\usetikzlibrary{arrows,automata,positioning}

\DeclareGraphicsExtensions{.pdf, .png, .jpg}
\graphicspath{{fig/}}

\tikzstyle{every state}=[inner sep=1pt, text width=1.5cm, align=center]
\tikzstyle{initial}=[text=, fill=black]
\tikzset{
block/.style={
  draw, 
  rectangle, 
  minimum height=1.5cm, 
  minimum width=2.1cm, align=center
  }, 
line/.style={->,>=latex'}
}
\tikzumlset{font=\scriptsize}
\newcommand*\circled[1]{\tikz[baseline=(char.base)]{
            \node[shape=circle,draw] (char) {#1};}}
\newcommand*\inlinecircled[1]{\tikz[yshift=2.65ex]{
            \node[shape=circle,draw,scale=0.8] (char) {#1};}}

\newcommand{\ie}{\emph{i.e.,}\xspace}
\newcommand{\eg}{\emph{e.g.,}\xspace}
\newcommand{\cf}{\emph{cf.}\xspace}

\newcommand{\etc}{\emph{etc.}\xspace}
\newcommand{\SOFTWARE}{LaTeXDraw\xspace}
\newcommand{\spoonVis}{\emph{Spoon Visualisation}\xspace}
\newcommand{\dltimes}{\SI{1.5}{k}\xspace}
\newcommand{\wb}{\emph{Interacto}\xspace}
\newcommand{\wblit}{Interacto binding\xspace}
\newcommand{\wber}{Interacto binder\xspace}
\newcommand{\wbers}{Interacto binders\xspace}
\newcommand{\wbslit}{Interacto bindings\xspace}
\newcommand{\ijfx}{\emph{Interacto-JavaFX}\xspace}
\newcommand{\iang}{\emph{Interacto-Angular}\xspace}
\newcommand{\its}{\emph{Interacto-TypeScript}\xspace}
\newcommand{\ijava}{\emph{Interacto-Java}\xspace}

\newcommand\pc[1]{\SI{#1}{\percent}\xspace}
\newcommand\s[1]{\SI{#1}{\second}\xspace}

\theoremstyle{definition}
\newtheorem{definition}{Definition}

\definecolor{kw}{rgb}{0.5,0.,0.33}
\definecolor{com}{rgb}{0.247,0.5,0.372}
\definecolor{str}{rgb}{0.165,0.0,1.0}
\definecolor{ano}{rgb}{0.4,0.4,0.4}
\definecolor{bg}{rgb}{1,1,1}
\definecolor{mygray}{rgb}{0.9,0.9,0.9}

\setabbreviationstyle[acronym]{long-short}
\newacronym{hci}{HCI}{Human-Computer Interaction}
\newacronym{mvc}{MVC}{Model-View-Controller}
\newacronym{ui}{UI}{User Interface}
\newacronym{gui}{GUI}{Graphical User Interfaces}
\newacronym{dnd}{DnD}{Drag-And-Drop}
\newacronym{fsm}{FSM}{Finite-State Machine}
\newacronym{rp}{RP}{Reactive Programming}
\newacronym{ast}{AST}{Abstract Syntax Tree}

\newcommand{\ui}{\gls{ui}\xspace}
\newcommand{\uis}{\glspl{ui}\xspace}

\lstdefinelanguage{myjs}{
  stringstyle=\color{str}, 
  keywords={typeof, new, true, false, catch, function, return, null, catch, switch, var, if, in, while, do, else, case, break, let, const},
  keywordstyle={\color{kw}\bfseries}, 
  commentstyle=\color{com}, 
  xleftmargin=0.5em, 
  xrightmargin=0.5em,
  basicstyle={\footnotesize\ttfamily}, 
  captionpos=b,  
  showstringspaces=false,
  frame=single,
  escapechar=|, 
  numbersep=5pt, 
  numbers=left, 
  tabsize=2,
  backgroundcolor=\color{bg}, 
  moredelim=[is][\textcolor{ano}]{\%\%}{\%\%},
  morestring=[b]',
  morestring=[b]"
}

\lstdefinelanguage{MyTS}{
  stringstyle=\color{str}, 
  keywordstyle={\color{kw}\bfseries}, 
  commentstyle=\color{com}, 
  basicstyle={\ttfamily\linespread{1.2}\footnotesize},
  backgroundcolor=\color{mygray},  
  captionpos=b, 
  xleftmargin=0.5em, 
  xrightmargin=0.5em, 
  escapechar=~, 
  keywords={private, class, export, implements, void, const, as, new, instanceof, this},
  showstringspaces=false,
  frame=single,
  numbersep=5pt, 
  tabsize=2,
  moredelim=[is][\textcolor{ano}]{\%\%}{\%\%}
}

\lstdefinelanguage{MyHTML}{
  stringstyle=\color{str}, 
  keywordstyle={\color{kw}\bfseries}, 
  commentstyle=\color{com}, 
  basicstyle={\ttfamily\linespread{1.2}\footnotesize},
  captionpos=b,  
  showstringspaces=false,
  frame=single,
  escapechar=|, 
  numbersep=5pt, 
  xleftmargin=0.5em, 
  xrightmargin=0.5em,
  language=html, 
  tabsize=2,
  moredelim=[is][\textcolor{ano}]{\%\%}{\%\%}
}

\lstdefinelanguage{MyJava}{
  language=java, 
  stringstyle=\color{str}, 
  keywordstyle={\color{kw}\bfseries}, 
  commentstyle=\color{com}, 
  basicstyle={\ttfamily\linespread{1.2}\footnotesize},
  morekeywords={var},
  backgroundcolor=\color{mygray},  
  captionpos=b,  
  showstringspaces=false,
  frame=single,
  xleftmargin=0.5em, 
  xrightmargin=0.5em,
  escapechar=|, 
  numbersep=5pt, 
  tabsize=2,
  moredelim=[is][\textcolor{ano}]{\%\%}{\%\%}
}

\lstdefinelanguage{MyJava2}{
  language=java, 
  stringstyle=\color{str}, 
  keywordstyle={\color{kw}\bfseries}, 
  commentstyle=\color{com}, 
  basicstyle={\ttfamily\footnotesize},
  morekeywords={var},
  xleftmargin=0.5em, 
  xrightmargin=0.5em,
  backgroundcolor=\color{bg},  
  captionpos=b,  
  showstringspaces=false,
  frame=single,
  escapechar=|, 
  numbersep=5pt, 
  tabsize=2,
  moredelim=[is][\textcolor{ano}]{\%\%}{\%\%}
}

\definecolor{linkColor}{RGB}{6,125,233}
\hypersetup{%
  pdftitle={},
  pdfauthor={},
  pdfkeywords={},
  pdfdisplaydoctitle=true,
  bookmarksnumbered,
  pdfstartview={FitH},
  colorlinks,
  citecolor=black,
  filecolor=black,
  linkcolor=black,
  urlcolor=linkColor,
  breaklinks=true,
  hypertexnames=false
}

\begin{document}

\title{Interacto: A Modern User Interaction Processing Model}

\author{Arnaud Blouin
         and Jean-Marc Jézéquel%
 \IEEEcompsocitemizethanks{\IEEEcompsocthanksitem M. Blouin and M. Jézéquel were with Univ Rennes, Inria, CNRS, IRISA, France. E-mail: firstname.lastname@irisa.fr}}

\IEEEtitleabstractindextext{%

Since most software systems provide their users with interactive features, building user interfaces (UI) is one of the core software engineering tasks. 
It consists in designing, implementing and testing ever more sophisticated and versatile ways for users to interact with software systems, and safely connecting these interactions with commands querying or modifying their state.
However, most UI frameworks still rely on a low level model, the bare bone UI event processing model. 
This model was suitable for the rather simple UIs of the early 80's (menus, buttons, keyboards, mouse clicks), but now exhibits major software engineering flaws for modern, highly interactive UIs. These flaws include lack of separation of concerns, weak modularity and thus low reusability of code for advanced interactions, as well as low testability.
To mitigate these flaws, we propose \wb as a high level user interaction processing model. By reifying the concept of user interaction, \wb makes it easy to design, implement and test modular and reusable advanced user interactions, and to connect them to commands with built-in undo/redo support.
To demonstrate its applicability and generality, we briefly present two open source implementations of \wb for Java/JavaFX and TypeScript/Angular. 
We evaluate \wb interest (1) on a real world case study, where it has been used since 2013, and with (2) a controlled experiment with 44 master students, comparing it with traditionnal UI frameworks.
 }

\maketitle

\section{Introduction}\label{sec:introduction}

"\emph{Anytime you turn on a computer, you're dealing with a user interface}"~\cite{6484050}.
\uis, and the user interactions they supply, pervade our daily lives by enabling users to interact with software systems.
The user interactions provided by a \ui form a dialect between a system and its users~\cite{Green86}:
a given user interaction can be viewed as a sentence composed of predefined words, i.e. low-level \ui events, such as mouse pressure or mouse move.
For example, we can view the execution of a drag-and-drop interaction as a sentence emitted by a user to the system.
This sentence is usually composed of the words \emph{pressure}, \emph{move}, and \emph{release}, that are \ui events assembled in this specific order.
The human-computer interaction community designs novel and complex user interactions.
As explained by \cite{beaudouin2004designing}, "\emph{\gls{hci} researchers have created a variety of novel [user] interaction techniques and shown their effectiveness in the lab [...]. 
Software developers need models, methods, and tools that allow them to transfer these techniques to commercial applications.}"
Currently, to use such novel user interactions in software systems developers must complete two software engineering tasks:
\begin{enumerate*}[label=(\roman*)]
\item They must assemble low-level \ui events to build the expected user interaction.
For example, a developer must manually assemble the events \emph{pressure}, \emph{move}, and \emph{release} to build a drag-and-drop.
\item They have to code how to process such \ui events when triggered by users.
\end{enumerate*}

\begin{figure*}[t]
\centering
\scalebox{0.9}{
\begin{tikzpicture}[->, thick]
   \node[block] (sw) {
      Software System\\[2mm]
      \begin{tikzpicture}[align=center]
         \node[block] (m) {Data Model};
         \node[block, left=1.5cm of m] (con) {
            Controllers\\[2mm]
            \begin{tikzpicture}[align=center]
               \node[block, minimum height=0.7cm, minimum width=0.7cm] (con1) {$cont_1$};
               \node[block, minimum height=0.7cm, minimum width=0.7cm, below=0.4cm of con1] (con2) {$cont_2$};
            \end{tikzpicture}
         };
         \node[block, left=1.5cm of con] (ui) {
            User Interface\\[2mm]
            \begin{tikzpicture}[align=center]
               \node[block, minimum height=0.7cm, minimum width=0.7cm] (w1) {$o_1$};
               \node[block, minimum height=0.7cm, minimum width=0.7cm, below right=0.4cm of w1] (w2) {$o_2$};
            \end{tikzpicture}
         };
      \end{tikzpicture}
   };
   \node[block, left=1.5cm of sw] (u) {User};
   \path(u) edge[above] node[align=center] {\small \circled{1}} (-4.5,0);
   \path(-3.75,0) edge[above] node[align=center] {\small \circled{2}} (-0.4,0);
   \path(0.55,0) edge[above] node[align=center] {\small \circled{3}} (2.65,0);
\end{tikzpicture}
}
\caption{Standard behavior of the \gls{ui} event processing model:\\[-0.05cm]
\protect\tikz \protect\inlinecircled{1}; \hspace*{0.2cm}: A user interacts with an interactive object $o_1$ of the user interface.\\[-0.05cm]
\protect\tikz \protect\inlinecircled{2}; \hspace*{0.2cm}: The interactive object then triggers a \gls{ui} event gathered by a controller $cont_1$.\\[-0.05cm]
\protect\tikz \protect\inlinecircled{3}; \hspace*{0.2cm}: The controller contains an event callback that processes this \gls{ui} event to possibly modify the business data.}
\label{fig.basicProcess}
\vspace*{-0.4cm}
\end{figure*}

To do so, developers still use a technique proposed with SmallTalk and the \emph{Model-View-Controller} (MVC) pattern in the 80's~\cite{krasner1988}:
the \ui event processing model, currently implemented using callback methods or reactive programming~\cite{bainomugisha2013} libraries.
This model considers low-level \ui events as the first-class concept developers can use for coding and using increasingly complex user interactions not supported off-the-shelf by graphical toolkits.
The reason is that interacting with classical widgets (\eg buttons, lists, menus) is usually one-shot:
a single \ui event, such as a mouse pressure on a button or menu, has to be processed.
For more complex user interactions such as the drag-and-drop, the current event processing model exhibits critical software engineering flaws that hinder code reuse and testability, and negatively affect separation of concerns and code complexity:
\begin{itemize}
   \item the concept of user interaction does not exist, so developers have to re-define user interactions for each \ui by re-coding them using \ui events;
   \item the model does not natively support advanced features, such as cancellation (undo/redo), event throttling, or logging;
   \item developers mix in the same code the assembly of \ui events and their processing, leading to a lack of separation of concerns;
   \item the use of callbacks to process \ui events
    (1) can lead to "spaghetti" code~\cite{Myers1991,oney2014interstate};
    (2) is based on the \emph{Observer} pattern that has several major drawbacks~\cite{maier2012deprecating,salvaneschi2014towards,Salvaneschi2014b,foust2015};
    (3) can be affected by design smells~\cite{blouin:hal-01499106};
\end{itemize}

This paper makes the following software engineering contribution:
a user interaction processing model called \wb that overcomes the above-mentioned flaws of the \ui event processing model.
\wb reifies user interactions and \ui commands as first-class objects and provides dedicated algorithms, object-oriented properties, run-time optimizations, and testing facilities to permit developers to stay focused on the core tasks of coding \uis:
\begin{enumerate*}[label=(\roman*)]
\item select the user interactions they have to use;
\item code how to turn these user interactions into undoable \ui commands;
\item reuse user interactions and \ui commands in different places across software systems;
\item write \ui tests.
\end{enumerate*}
In this model, \ui events are now considered as low-level implementation concepts rarely used by developers.

To demonstrate its applicability and generality, we developed two implementations of \wb: \ijfx with Java and JavaFX~\cite{jfx}, a mainstream Java graphical toolkit; \iang with TypeScript~\cite{bierman14} and Angular~\cite{angular}, a mainstream Web graphical toolkits.
Both implementations take the form of a fluent API (\emph{Application Programming Interface})~\cite{fowler2010}.
We evaluate \wb interest:
\begin{itemize}
   \item on a real world case study: the development \SOFTWARE, an open-source highly interactive vector drawing editor for \LaTeX{}, downloaded \dltimes times per month and available on more than \num{10} Linux distributions. 
   \item with a controlled experiment with 44 master students, comparing it with traditional UI frameworks.
\end{itemize}

The paper is organized as follows.
\Cref{sec.backMotiv} introduces the background concepts and motivates the work by detailing the limits of the current \ui event processing model.
\Cref{sec.approach} details the proposed user interaction processing model and its testing support.
\Cref{sec.eval} evaluates the proposal.
\Cref{sec.related} discusses the related research work.
\Cref{sec.conclusion} concludes the paper and discusses future work.

\section{Background and Motivations}\label{sec.backMotiv}

\subsection{Definitions}\label{sec.defs}

The standard \ui event processing model involves the following concepts, as depicted by \Cref{fig.basicProcess}.
\smallskip

\noindent \textbf{User Interface.} A \gls{ui} allows users to control or query a software system.
The most common kind of user interfaces are \glspl{gui}.
A \gls{ui} is composed of interactive objects, such as buttons or canvases for \glspl{gui}.

\smallskip
\noindent \textbf{\gls{ui} event.} When a user interacts with an interactive object, this last triggers a \emph{\gls{ui} event} such as \emph{mouse pressed} or \emph{key released}.
A \gls{ui} event embeds data, such as the position of the mouse pressure.

\smallskip
\noindent \textbf{\gls{ui} controller.}\footnote{For simplicity, we use the term \emph{controller} to refer to any kind of components that processes \gls{ui} events, such as \emph{Presenter} (MVP)~\cite{potel1996}, \emph{ViewModel} (MVVM~\cite{mvvm}), or \emph{Component} (Angular~\cite{angular}).} A \emph{controller} registers with different interactive objects to gather the \gls{ui} events these objects trigger.

\smallskip
\noindent \textbf{Event callback}. An event callback is a method associated to an interactive object.
The interactive object calls such a method on \gls{ui} event triggering.
Developers define event callbacks in controllers.
The goal of such callbacks is usually to modify the business data (but can be used to modify the state of the \gls{ui} as well).

\smallskip

In addition to the concepts involved in the event process model, we define the concepts of user interaction and \gls{ui} command as follows.
\smallskip

\noindent \textbf{User Interaction.} 
Users perform user interactions on a \gls{ui} to control the underlying system.
User interactions technically rely on one or a sequence of \gls{ui} events.
For example, a \gls{dnd} is the sequence of one pressure event, one or several move events, and one release event.
A user interaction is independent of its possible usages.
For example, one can use a \gls{dnd} for moving, scaling, or deleting objects.

\smallskip

\noindent \textbf{\gls{ui} command.} 
A user performs a user interaction on a \gls{ui} to apply a specific \gls{ui} command on the underlying system.
Examples of \gls{ui} commands applied using a \gls{dnd} are moving, scaling, or deleting objects. 
\gls{ui} command can take two shapes:
using callbacks~\cite{blouin:hal-01499106} such as in \Cref{lst.jsEx} discussed in the next section;
or using classes as discussed in \Cref{sec.related}.

\subsection{Motivating Example}\label{sec.example}

\Cref{lst.jsEx} contains JavaScript code, adapted from~\cite{oney2014interstate}, that illustrates the \ui event processing model depicted by \Cref{fig.basicProcess}.
In this code example, a user can move a graphical rectangular node using a drag-lock interaction.
During this drag-lock, the user interface uses a 'hand' cursor as user feedback.
The JavaScript code of \Cref{lst.jsEx} contains:
the coding of a drag-lock user interaction;
the use of this drag-lock interaction to move a graphical node.

The drag-lock user interaction is a special kind of drag-and-drop.
A drag-lock starts by double-clicking on the node to drag.
The user can then move the locked node until she double-clicks again at the dropping location.
The drag-lock interaction is an interesting motivating example as it is a standard user interaction but not provided off-the-shelf by the current \gls{ui} toolkits.

\noindent\begin{minipage}{\linewidth}
\begin{lstlisting}[language=myjs, numbers=left, xleftmargin=2em, label=lst.jsEx, caption={A JavaScript code snippet to move a node using a drag-lock, adapted from~\cite{oney2014interstate}}]
let isDragLocked = false;
const moveCallback = evt => {|\label{code.exjs10}|
	draggable.attr({ x: evt.x, y: evt.y });|\label{code.exjs1}|
};
draggable.addEventListener('dblclick', evt => {|\label{code.exjs7}|
	if (evt.button === 0) {|\label{code.exjs11}|
		if (isDragLocked) {|\label{code.exjs5}|
			draggable.style.cursor = '';
			draggable
			.removeEventListener('mousemove', moveCallback);|\label{code.exjs0}|
		} else {|\label{code.exjs4}|
			draggable.style.cursor = 'hand';
			draggable
			.addEventListener('mousemove', moveCallback);|\label{code.exjs2}|
		}
		isDragLocked = !isDragLocked;
	}
});|\label{code.exjs3}|
\end{lstlisting}
\end{minipage}

The drag-lock of \Cref{lst.jsEx} assembles the \ui events \emph{dblclick} (double-click) and \emph{mousemove}.
\Cref{code.exjs7}, the node to drag (\emph{draggable}) registers to double click events.
The second argument of this function is a callback method executed on each double-click on this node (\Crefrange{code.exjs7}{code.exjs3}).
For the first double-click, the \ui uses the 'hand' cursor and the node registers to mouse move events (\Crefrange{code.exjs4}{code.exjs2}).
For the second double-click, the \ui uses the default cursor and the node unregisters to mouse move events (\Crefrange{code.exjs5}{code.exjs0}).
The (un-)registration to mouse move events takes as second argument the callback method located \Cref{code.exjs10} that moves the node using event data (\Cref{code.exjs1}).
The move of the node operates only if the user uses the mouse button $0$ (\Cref{code.exjs11}).

This code mixes both the definition of the user interaction and its use for moving a node.
Moreover, coding a user interaction may require coding specific instructions for manually registering and unregistering to \ui events (\Cref{code.exjs0,code.exjs2}).

\vspace*{-0.8cm}
\begin{figure}[h]
	\centering
	\begin{tikzpicture}[->, node distance=3cm, thick]
		\node[initial,state, minimum size=1cm, inner sep=0pt, text width=0cm](Init){};
		\node[state, minimum size=1.4cm](Locked)[right of=Init]{{\small Locked}};
		\node[state, minimum size=1.4cm, accepting](Unlocked)[right of=Locked]{{\small Unlocked}};
		\path(Init) edge[above] node[align=center] {\small double\\click} (Locked);
		\path(Locked) edge[loop, right, in=20,out=60,looseness=6] node {{\small move}} (Locked);
		\path(Locked) edge[below] node[align=center] {\small double\\click} (Unlocked);
	\end{tikzpicture}
	\smallskip
	
	\begin{tikzpicture}[->, node distance=3cm, thick]
		\node[initial,state, minimum size=1cm, inner sep=0pt, text width=0cm](Init){};
		\node[state, minimum size=1.4cm](Clicked)[right of=Init]{{\small Clicked}};
		\node[state, minimum size=1.4cm, accepting](DbleClicked)[right of=Clicked]{{\small Double Clicked}};
		\node[state, minimum size=1.4cm, accepting, yshift=-1.2cm](Cancelled)[above of=DbleClicked]{{\small Cancelled}};
		\path(Init) edge[above] node[align=center] {\small click} (Clicked);
		\path(Clicked) edge[bend left] node[align=center, xshift=-1.3cm] {\small timeout $[t\ge 1s]$} (Cancelled);
		\path(Clicked) edge[above] node[align=center] {\small click} (DbleClicked);
	\end{tikzpicture}
	\caption{\glspl{fsm} of the drag-lock (top) and double-click (bottom) user interactions used in \Cref{lst.jsEx}. The double-click transition used in the drag-lock \gls{fsm} refers to the double-click interaction.}
	\label{fig.dragLock}
\end{figure}
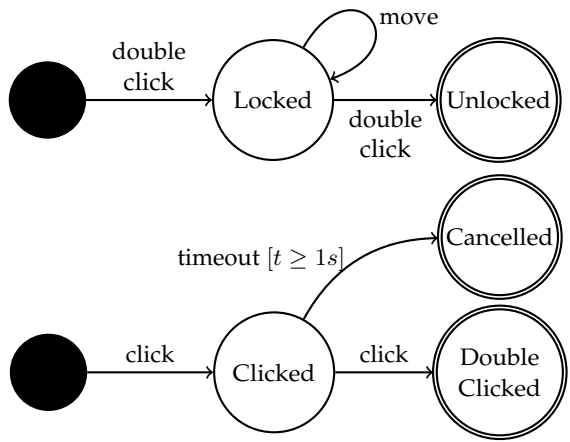

\Cref{fig.dragLock} (on the top) depicts an \gls{fsm} that illustrates the assembly of the \emph{mousemove} and \emph{dblclick} events to build the drag-lock.
A transition refers to a \gls{ui} event or another user interaction.
The execution of one user interaction ends when its \gls{fsm} reaches a terminal state.
One may notice that some \gls{ui} events are not atomic:
if the double-click is a user interaction based on several events (\emph{pressure} and \emph{release} that compose each click), it is sometimes considered as a \ui event since it is one-shot.
Also using an \gls{fsm}, \Cref{fig.dragLock} (on the bottom) depicts the assembly of \gls{ui} events to build a standard double-click.

\subsection{Limitations of the UI event processing model}\label{sec.motiv}

We illustrate the current limitations of \gls{ui} event processing models using the example introduced in the previous section and depicted by \Cref{lst.jsEx}.
This example, that involves a drag-lock user interaction, suffers of the following flaws:

\medskip
\noindent\textbf{Lack of separation of concerns.}
\Cref{lst.jsEx} illustrates how relying on \gls{ui} events breaks the concept of separation of concerns~\cite{parnas1972criteria} by intertwining in the same code:
\begin{itemize}\itemsep0cm
   \item The definition of the user interaction (the drag-lock) that consists of the assembly of \gls{ui} events.
   Current \gls{ui} toolkits and approaches consider \gls{ui} events as a first-class concept for coding user interfaces.
   \gls{ui} events, however, are \emph{low-level implementation details} that developers need to manually assemble to build user interactions, such as the drag-lock.

   \item The transformation of user interactions into \gls{ui} commands.
   In the same code that assembles \gls{ui} events to build a user interaction, developers have to define how to produce output \gls{ui} commands.
   \Cref{code.exjs1} in \Cref{lst.jsEx} is the command instruction that moves the dragged node.
   
   \item Conditions that constraint the execution of the user interaction.
   For example, \Cref{code.exjs11} checks whether the drag-lock has been done using the button $0$ of the mouse.
\end{itemize}

\medskip
\noindent\textbf{Lack of software reuse.}
\Cref{lst.jsEx} also illustrates how the \gls{ui} event processing model prevents code reuse~\cite{Krueger1992,Johnson1997}:
\begin{itemize}\itemsep0cm
   \item \textbf{No user interaction reuse.} 
   Libraries and frameworks enable software reuse by providing developers with predefined and reusable artifacts.
   Ignoring the concept of user interactions prevents the development of reusable interactions based on the designs established by the \gls{hci} community.

   \item \textbf{No user interaction substitution}.
   User interactions can be classified in different categories.
For example, the drag-lock is a kind of \gls{dnd} interactions.
Following the object-oriented substitutability concept, a developer should easily be able to replace a \gls{dnd} with a drag-lock as their underlying data are the same:
start and end positions.
Moreover, a same user interaction may have behavioral variants.
\Cref{fig.dragLockAlt} depicts alternative behaviors of the drag-lock and double-click interactions.
The double-click is now canceled on a move between the two clicks.
The timeout has changed to \SI{0.5}{\second}.
The drag-lock now requires at least one move between the two double-clicks, otherwise it is canceled.
A pressure on the key 'ESC' cancels the user interaction.
In such cases of user interaction variants, a developer should easily  be able to replace the standard \gls{dnd} by a variant, still based on substitutability.

Developers can hardly achieve user interaction substitution with the current \gls{ui} event processing model as the assembly of \gls{ui} events has to be modified and this model lacks object-oriented constructs.

\begin{figure}[h]
\centering
\scalebox{0.9}{
\begin{tikzpicture}[->, node distance=2.5cm, thick]
\node[initial,state, minimum size=1cm, inner sep=0pt, text width=0cm](Init){};
\node[state, minimum size=1.4cm](Locked)[right of=Init]{{\small Locked}};
\node[state, minimum size=1.4cm](Moved)[right of=Locked]{{\small Moved}};
\node[state, minimum size=1.4cm, accepting](Unlocked)[right of=Moved]{{\small Unlocked}};
\node[state, minimum size=1.4cm, accepting](Canceled)[above of=Moved]{{\small Canceled}};
\path(Init) edge[above] node[align=center] {\small double\\click} (Locked);
\path(Locked) edge[below] node {{\small move}} (Moved);
\path(Locked) edge[bend angle=40, bend left] node[above left] {{\small key press [key='ESC']}} (Canceled);
\path(Locked) edge[bend angle=25, bend right] node[above left, align=center] {\small double\\click} (Canceled);
\path(Moved) edge[above right] node {{\small key press [key='ESC']}} (Canceled);
\path(Moved) edge[loop, right, in=20,out=60,looseness=6] node {{\small move}} (Moved);
\path(Moved) edge[below] node[align=center] {\small double\\click} (Unlocked);
\end{tikzpicture}
}
\smallskip

\scalebox{0.9}{
\begin{tikzpicture}[->, node distance=2.2cm, thick]
\node[initial,state, minimum size=1cm, inner sep=0pt, text width=0cm](Init){};
\node[state, minimum size=1.4cm](Clicked)[right of=Init]{{\small Clicked}};
\node[state, minimum size=1.4cm, accepting](DbleClicked)[below of=Clicked]{{\small Double Clicked}};
\node[state, minimum size=1.4cm, accepting](Canceled)[right of=Clicked]{{\small Canceled}};
\path(Init) edge[above] node[align=center] {\small click} (Clicked);
\path(Clicked) edge[bend right] node[align=center] {\small ~\\~\\~\\timeout\\$[t\ge 0.5s]$} (Canceled);
\path(Clicked) edge[bend left] node[align=center] {\small move\\~} (Canceled);
\path(Clicked) edge[left] node[align=center] {\small click} (DbleClicked);
\end{tikzpicture}
}
\caption{Alternative versions of the drag-lock (top) and the double-click (bottom) user interactions}
\label{fig.dragLockAlt}
\end{figure}
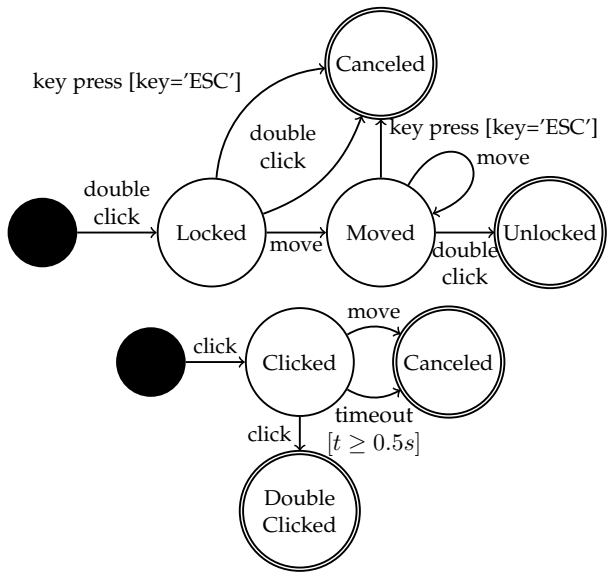
\end{itemize}

\medskip
\noindent\textbf{Lack of advanced features.}
\begin{itemize}
   \item \textbf{No undo/redo.} 
   The code of \Cref{lst.jsEx} modifies the business data directly in the event callbacks (\Cref{code.exjs1}).
   So, the changes cannot be stored to be then undone and redone.
   This would require glue code manually crafted by developers in the code of \Cref{lst.jsEx} to support such a feature.
   Note that several UI toolkits overcome this lack with dedicated features (see \Cref{sec.related}).

   \item \textbf{No logging.}
   Modern systems use logs for analyzing \gls{ui} usages~\cite{Apaolaza:2017} and for understanding issues.
   The \gls{ui} event processing model does not support logging natively.
   
   \item \textbf{No throttling.}
   Event throttling is an optimization that permits to reduce the number of similar and successive events in order to alleviate the processing load (and possibly to gain performance).
   The \gls{ui} event processing model does not support such a feature.
\end{itemize}

\medskip
\noindent\textbf{Complexity and design issues.}
Intertwining in the same code the assembly of \gls{ui} events to build user interactions and the transformation of \gls{ui} events into commands makes the code more complex:
it can lead to "spaghetti" code~\cite{Myers1991,oney2014interstate} and can be affected by design smells~\cite{blouin:hal-01499106}.
Moreover, the \gls{ui} process model strongly relies on the \emph{Observer} pattern that suffers from several major flaws~\cite{maier2012deprecating,salvaneschi2014towards,Salvaneschi2014b,foust2015}.

This also makes the code \emph{more difficult to test}.
For example with the code of \Cref{lst.jsEx}, developers have to test the assembly of \ui events.

\section{The user interaction processing model}\label{sec.approach}

This section describes a user interaction processing model we named \wb, that overcomes the limitations detailed in the previous section.
User interactions form a core concept of this model instead of events.
So, we call this model a user interaction processing model instead of an event processing model.
\textbf{The gist of \wb is to turn user interaction executions into (undoable) commands.}

\begin{definition}
\textbf{\wblit.}
An \wblit is an object that turns the executions of one user interaction into (undoable) command instances.
\end{definition}

\begin{definition}
	\textbf{\wber.}
An \wber is an object that configures one specific \wblit.
In the \wb implementations, an \wber takes the form of a fluent API.
\end{definition}

\Cref{sub.core} gives an overview of the \wb approach.
\Cref{sub.interaction,sub.cmd} describes how user interactions and \ui commands work in \wb.
\Cref{sub.model,sub.syntax} then focus on the \wblit behavior and the \wber syntax.
\Cref{sub.test} details the benefits of \wb in terms of \ui testing by proposing new UI testing oracles.
Finally, \Cref{sub.props} describes properties that characterize \wbers.

All the examples of this section are based on our Java implementation of \wb, namely \ijfx.
The use of our TypeScript implementation, \iang, would have led to very similar code examples.

\subsection{Approach overview}\label{sub.core}

\begin{figure*}[h]
\centering
\scalebox{0.85}{
\begin{tikzpicture}[->, thick]
   \node[block] (sw) {
      Software System\\[2mm]
      \begin{tikzpicture}[align=center]
         \node[block] (m) {Data Model};
         \node[block, left=0.6cm of m] (cmd) {
            \textbf{Commands}\\[2mm]
            \begin{tikzpicture}[align=center]
            \node[block, minimum height=0.7cm, minimum width=0.7cm] (c1) {$cmd_1$};
            \node[block, minimum height=0.7cm, minimum width=0.7cm, below=0.4cm of c1] (c2) {$cmd_2$};
            \end{tikzpicture}
         };
         \node[block, left=0.6cm of cmd] (con) {
            Controllers\\[2mm]
            \begin{tikzpicture}[align=center]
               \node[block, minimum height=0.7cm, minimum width=0.7cm] (con1) {$cont_1$};
               \node[block, minimum height=0.7cm, minimum width=0.7cm, below=0.4cm of con1] (con2) {$cont_2$};
            \end{tikzpicture}
         };
         \node[block, left=0.6cm of con] (inter) {
            \textbf{User interactions}\\[2mm]
            \begin{tikzpicture}[align=center]
               \node[block, minimum height=0.7cm, minimum width=0.7cm] (inter1) {$inter_1$};
               \node[block, minimum height=0.7cm, minimum width=0.7cm, below=0.4cm of con1] (inter2) {$inter_2$};
            \end{tikzpicture}
         };
         \node[block, left=0.6cm of inter] (ui) {
            User Interface\\[2mm]
            \begin{tikzpicture}[align=center]
               \node[block, minimum height=0.7cm, minimum width=0.7cm] (w1) {$o_1$};
               \node[block, minimum height=0.7cm, minimum width=0.7cm, below right=0.4cm of w1] (w2) {$o_2$};
            \end{tikzpicture}
         };
      \end{tikzpicture}
   };
   \node[block, left=0.8cm of sw] (u) {User};
   \path(u) edge[above] node[align=center] {\small \circled{1}} (-6.6,0);
   \path(-5.9,0) edge[above] node[align=center] {\small \circled{\textbf{2}}} (-3.2,0);
   \path(-2.1,0) edge[above] node[align=center] {\small \circled{\textbf{3}}} (-0.2,0);
   \path(0.9,0) edge[above] node[align=center] {\small \circled{\textbf{4}}} (2.6,0);
   \path(3.6,0) edge[above] node[align=center] {\small \circled{\textbf{5}}} (5,0);
\end{tikzpicture}
}
\caption{Behavior of the proposed user interaction processing model (in bold what differs from \Cref{fig.basicProcess}):\\
\protect\tikz \protect\inlinecircled{1}; \hspace*{0.2cm}: A user interacts with an interactive object $o_1$ of the user interface.\\
\protect\tikz \protect\inlinecircled{2}; \hspace*{0.2cm}: The interactive object then triggers a \ui event \textbf{processed by a running user interaction $inter_1$}.\\
\protect\tikz \protect\inlinecircled{3} and \protect\inlinecircled{4}; \hspace*{0.6cm}: \textbf{Controllers contain \wbslit that turns user interaction executions into \ui commands}.\\
\protect\tikz \protect\inlinecircled{5}; \hspace*{0.2cm}: \textbf{The running \wblit executes the ongoing \ui command} to modify the state of the system.}
\label{fig.bindingProcess}
\vspace*{-0.2cm}
\end{figure*}
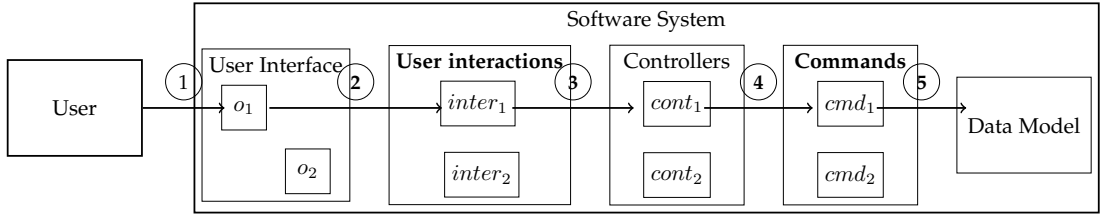

Consider the example of \Cref{sec.example}:
users have to use a drag-lock interaction to translate a graphical object.
\Cref{lst.pseudo} illustrates how \wb works in pseudo-code.
The developer: 
selects one user interaction (\Cref{code.pseudo1});
specifies the widgets on which the \wblit will operate (\Cref{code.pseudo2});
selects the command to produce (\Cref{code.pseudo3});
details what to do during the execution of the user interaction, in particular when the interaction starts, updates, and ends or is canceled (\Crefrange{code.pseudo4}{code.pseudo5});
defines the conditions that constraint the production and the execution of the ongoing command (\Cref{code.pseudo6}).

\medskip
\begin{lstlisting}[xleftmargin=0.6em,   basicstyle={\ttfamily\linespread{1.2}\footnotesize}, captionpos=b, numbers=left, xleftmargin=1.8em, showstringspaces=false, frame=single, escapechar=|, numbersep=5pt, caption={Pseudo-code of an \wber that configures an \wblit that translates a node using a drag-lock}, label={lst.pseudo}]
Use the drag-lock user interaction,|\label{code.pseudo1}|
On the interactive object 'node',|\label{code.pseudo2}|
To produce and execute 'Translate' command instances,|\label{code.pseudo3}|
When the interaction starts, sets a shadow to 'node',|\label{code.pseudo4}|
When the interaction stops or is cancelled, removes
     this shadow,
When the interaction updates, updates the translation
     vector that the ongoing command will use,|\label{code.pseudo5}|
That, only if the user uses the primary mouse button.|\label{code.pseudo6}|
\end{lstlisting}

This pseudo-code illustrates the four key concepts, that \Cref{fig.bindingProcess} depicts, on which \wb relies.
These concepts, detailed in the next sub-sections, are:
\begin{itemize}
   \item User interactions are reusable, composable (one can build a user interaction using other user interactions), and stateful objects that graphical libraries should provide to developers instead of low-level \ui events.

   \item \ui commands are reusable and undoable objects.

   \item An \wblit transforms executions of one user interaction into output \ui (undoable) commands.
  
  \item An \wber has properties and a concrete syntax for configuring \wbslit.
\end{itemize}

\subsection{User interaction}\label{sub.interaction}

A user interaction is composed of two separated elements:
its \emph{behavior} and its \emph{data}.

\smallskip
\noindent\textbf{Interaction behavior.} The proposed model makes no assumption on how the behavior of a user interaction is defined.
This can be, for example, using \glspl{fsm}~\cite{appert2008,blouin:inria-00477627}, Petri nets~\cite{NAV09} or reactive programming~\cite{czaplicki2013}.
Our implementation and the description of the approach make use of \glspl{fsm}.
We already detailed how we model user interactions using \glspl{fsm} in \Cref{sec.example} by discussing the examples of \Cref{fig.dragLock,fig.dragLockAlt}.

\vspace*{-0.4cm}
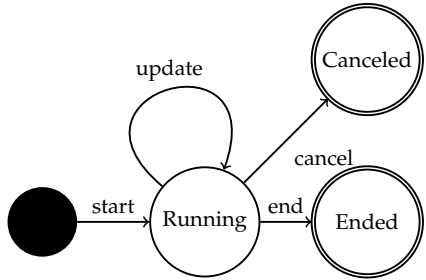
\begin{figure}[h]
   \centering
   \scalebox{0.9}{
      \begin{tikzpicture}[->, node distance=2.4cm, thick]
         \node[initial,state, minimum size=1cm, inner sep=0pt, text width=0cm](Init){};
         \node[state, minimum size=1.5cm](Running)[right of=Init]{{\small Running}};
         \node[state, minimum size=1.5cm, accepting](Ended)[right of=Running]{{\small Ended}};
         \node[state, minimum size=1.5cm, accepting](Canceled)[above of=Ended]{{\small Canceled}};
         \path(Init) edge[above] node[align=center] {\small start} (Running);
         \path(Running) edge[above] node[align=center] {\small end} (Ended);
         \path(Running) edge[above] node[below right, align=center] {\small cancel} (Canceled);
         \path(Running) edge[loop, right, in=70,out=140,looseness=6] node[above] {{\small update}} (Running);
      \end{tikzpicture}
   }
   \caption{User interaction life cycle}\label{fig.lifeCycleInter}
\end{figure}

\Cref{fig.lifeCycleInter} depicts the generic life cycle of any user interaction.
An interaction starts when its \gls{fsm} leaves its initial state, and is then considered as running.
A transition of an interaction \gls{fsm} corresponds to a \gls{ui} event or to another interaction \gls{fsm} (composite FSM).
A transition is executed when its matching \gls{ui} event is triggered (or when its sub-interaction has ended).
Each time a transition of the interaction \gls{fsm} is executed and the targeted state is not a terminal state, the interaction is updated:
user interactions update their interaction data (based on the data of the \gls{ui} event) on transitions executions.
Moreover, user interactions automatically perform optimizations on entry and exit actions of states (see \Cref{eval.implem}).
An interaction can end in two ways.
Either the interaction is canceled (state \emph{Canceled}), \ie the user wants to abort the interaction not to produce a command.
Either the interaction ends normally (state \emph{Ended}), \ie the user completed the user interaction.
We specify these two different terminal states in interaction \gls{fsm} examples (\Cref{fig.dragLock,fig.dragLockAlt}) by naming a canceling state \emph{Canceled}.
For example in \Cref{fig.dragLock} on the right, the \gls{fsm} reaches the state \emph{Canceled} when a timeout of 1 second expires.
The data of a user interaction are updated at each step (\ie on each transition execution) of their life cycle.

We define an \textbf{interaction execution} $exec_i$ as a path in the interaction life cycle, \ie a path: $start \rightarrow \ldots \rightarrow (cancel | end)$.
Note that an interaction execution also corresponds to a path of the \gls{fsm} of the interaction.
For example with the \gls{fsm} of the drag-lock interaction of \Cref{fig.dragLockAlt}, the \gls{fsm} path: $double~click \rightarrow move \rightarrow move \rightarrow double~click$, corresponds to the following path in the interaction life cycle: $start \rightarrow update \rightarrow update \rightarrow end$.

\begin{figure}[h]
\centering
\begin{tikzpicture}
   \umlsimpleclass[x=-3,y=-2.6]{DragLock}{}{}
   \umlsimpleclass[x=-1.8,y=-3.4]{DragAndDrop}{}{}
   \umlsimpleclass[x=0,y=-4.2]{DragAndPick}{}{}
   \umlsimpleclass[x=1.8,y=-3.4]{DragAndPop}{}{}
   \umlsimpleclass[x=3,y=-2.6]{DwellAndSpring}{}{}
   \umlsimpleclass[x=0,y=-2.3,type=abstract]{DnDBase}{}{}
   \umlinterface[x=0,y=0]{FromToData}{
      \umlvirt{+ getSrcObject() : Node}\\
      \umlvirt{+ getTgtObject() : Node}\\
      \umlvirt{+ getSrcPosition(): Point}\\
      \umlvirt{+ getTgtPosition() : Point}\\
      \umlvirt{+ getButton() : Optional<Button>}
   }{}
   \umlunicompo[mult=1,arg=interactionData]{DnDBase}{FromToData}
   \umlinherit{DragLock}{DnDBase} 
   \umlinherit{DragAndDrop}{DnDBase}
   \umlinherit{DragAndPick}{DnDBase}
   \umlinherit{DwellAndSpring}{DnDBase}
   \umlinherit{DragAndPop}{DnDBase}
\end{tikzpicture}
\caption{The data model of user interactions that operate from a source position to a target position (\eg \gls{dnd} and drag-lock)}\label{fig.interData}
\end{figure}
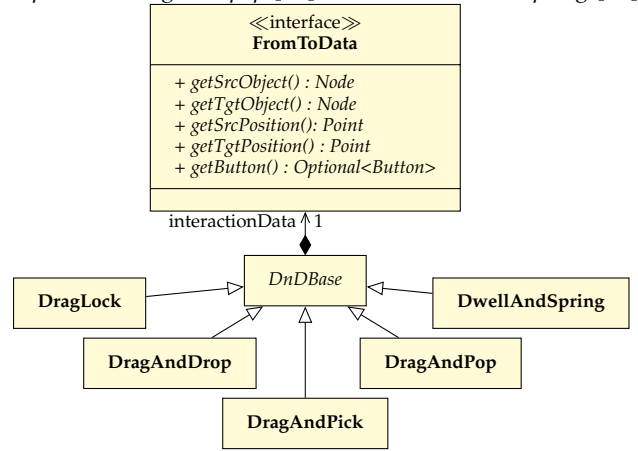

\smallskip
\noindent\textbf{Interaction data.}
A user interaction is stateful and exposes data that an \wblit can use.
The class diagram of \Cref{fig.interData} depicts the data model shared by both the drag-lock and the \gls{dnd} interactions.
The drag-lock and \gls{dnd} interactions are of the same type:
they consist in user interactions that operate from a source position to a target position.
The interface \emph{FromToData} represents the data of such user interactions, composed of:
the source position (\emph{getSrcPosition});
the source picked object (\emph{getSrcObject});
the target position (\emph{getTgtPosition});
the target picked object (\emph{getTgtObject});
the possible button if the interaction involves a mouse (\emph{getButton}).
Other user interactions complete this family, such as the \emph{drag-and-pick}, the \emph{drag-and-pop}~\cite{baudisch2003} and the \emph{dwell-and-spring}~\cite{appert2012}.

\vspace*{-0.4cm}

\noindent\textbf{Interaction type and substitution.}
During the definition of an \wblit, developers access the data of the selected user interaction, not its behavior.
For example with \Cref{fig.interData}, all the user interactions of the same family as the DnD expose the same type of interaction data \emph{FromToData}.
This choice follows the same reasoning than the \emph{bridge} design pattern~\cite{GAM95} that promotes the decoupling of interfaces and implementations:
user interaction exposes stable interaction data interfaces.
Developers base their configuration of an \wblit on such user interaction data so that this permits the substitution of user interactions of a same family.

Formally, we call an interaction data type $\mathcal{D}$ the interface that a user interaction $i$ of type $\mathcal{I}$ exposes.
This allows user interactions substitution:
developers can replace one user interaction $i$ of data type $\mathcal{D}$ (\eg a \gls{dnd} that exposes interaction data of type \emph{FromToData}), with another user interaction $i'$ of the same data type $\mathcal{D}$ (\eg a drag-lock, which interaction data type is also \emph{FromToData}) without any other change in the \wber.

\subsection{Undoable UI command}\label{sub.cmd}

\noindent\textbf{UI command.}
When a user interacts with a user interface using a given user interaction, her goal is to act on the underlying system, such as to modify its state.
Developers can encapsulate such actions on the system in \ui command classes.
In the literature, this has two benefits:
\begin{itemize}
   \item Enable \ui commands reuse across the different controllers of the \ui. One \ui command may be produced from different user interactions and interactive objects of the \ui (\eg buttons, menus, shortcuts).
   \item Support undo/redo features. developers may code \ui command as undoable, so that users can undo and redo changes they apply on the system.
\end{itemize}

\noindent\begin{minipage}{\linewidth}
\begin{lstlisting}[language=MyJava, numbers=left, xleftmargin=2em, caption={An example of an undoable UI command}, label={lst.uiCmd}, basicstyle={\ttfamily\linespread{1.1}\scriptsize}]
public class Translate extends CommandBase implements Undoable {
	double mementoX;
	double mementoY;
	double newX;|\label{code.uiCmd6}|
	double newY;|\label{code.uiCmd7}|
	final Shape data;|\label{code.uiCmd5}|
	
	public Translate(Shape shape) {
		data = shape;
	}
	@Override public boolean isExecutable() {|\label{code.uiCmd2}|
		return !(newX == data.getX() && newY == data.getY());
	}
	@Override protected void execution() {|\label{code.uiCmd1}|
		data.setPosition(newX, newY);
	}
	@Override protected void createMemento() {|\label{code.uiCmd8}|
		mementoX = data.getX();
		mementoY = data.getY();
	}
	@Override public void undo() {|\label{code.uiCmd3}|
		data.setPosition(mementoX, mementoY);
	}
	@Override public void redo() {|\label{code.uiCmd4}|
		execution();
}}
\end{lstlisting}
\end{minipage}

\Cref{lst.uiCmd}\footnote{For readability, all the code that uses an implementation of \wb is put in listings with a grayed background.} gives the Java code of a typical \wb \ui command coded by a developer (where \emph{CommandBase} and \emph{Undoable} are part of \wb).
The pseudo-code of \Cref{lst.pseudo} uses this \ui command.
This \ui command \emph{Translate} moves the given shape (\Cref{code.uiCmd5}) to a new position (\Cref{code.uiCmd6,code.uiCmd7}).
The method \emph{execution} defines the execution of the command (\Cref{code.uiCmd1}).
The \emph{isExecutable} method checks whether the \ui command can be executed (\Cref{code.uiCmd2}).
Here, this method checks that the translation vector is not null.
This command is undoable (its implements the interface \emph{Undoable}) so that the developer has to implement the methods \emph{undo} and \emph{redo} (\Cref{code.uiCmd3,code.uiCmd4}).
The \emph{undo} method puts the shape back to its former location using a memento~\cite{GAM95} produced by the \emph{createMemento} method (\Cref{code.uiCmd8}).
\wbslit automatically store the executed undoable commands in an undoable command history.
\wb supplies developers with predefined \ui commands for helping developers in adding undo/redo features in their \uis.

\medskip
\noindent\textbf{Asynchronous command.}
A developer may want to run a command asynchronously.
For example, a Web application may send queries to servers.
In such a case, the \emph{execution} method of an \wb command can return a Promise object that corresponds to the pending query.
In TypeScript, such a method can be written as follows:

\begin{lstlisting}[language=MyTS]
protected execution(): Promise<void> | void {
    return this.http.put(...).toPromise();
}
\end{lstlisting}

\medskip
\noindent\textbf{Undo history.}
An undo history collects the executed undoable commands and implements a specific undo mechanism.
In the literature, the standard undo mechanism is linear:
a new command is pushed on a stack \emph{toUndo};
on undo, the top command of this stack is popped to be undone and pushed on a second stack \emph{toRedo};
the opposite process operates on redo.

The literature proposed several undo mechanisms such as selective ones~\cite{prakash1994,yoon2015}.
\wb does not constraint the use of specific undo mechanisms and by default provides a linear undo history.
\wb also permits to have several undo histories in the same application to work with given \wblit as detailed in the next section.

\subsection{\wblit behavior}\label{sub.model}

\begin{figure}[h]
\centering
\begin{tikzpicture}
\umlsimpleclass[x=-2.3,y=-1.5,type=abstract]{Undoable}{}{}
\umlsimpleclass[x=0,y=1.3]{InteractoContext}{}{}
\umlsimpleclass[x=-2,y=0]{UndoRedoHistory}{}{}
\umlsimpleclass[x=1,y=-3,type=abstract]{InteractionBehavior}{}{}
\umlsimpleclass[x=-1.5,y=-3,type=abstract]{InteractionData}{}{}
\umlsimpleclass[x=2,y=-1.5,type=abstract]{Command}{}{}
\umlsimpleclass[x=-0.1,y=-1.5,type=abstract]{UserInteraction}{}{}
\umlsimpleclass[x=1,y=0]{InteractoBinding}{}{}
\umlunicompo[mult=1,arg=interaction,anchor1=-164,anchor2=100]{InteractoBinding}{UserInteraction}
\umlunicompo[mult=0..1,arg=currentCommand,anchor1=-26,anchor2=130]{InteractoBinding}{Command}
\umlunicompo[mult=1,arg=data,anchor1=-157,anchor2=30]{UserInteraction}{InteractionData}
\umlunicompo[mult=1,arg=behavior,anchor1=-20,anchor2=110]{UserInteraction}{InteractionBehavior}
\umlunicompo[mult=0..*,arg=bindings,anchor1=-20,anchor2=100]{InteractoContext}{InteractoBinding}
\umlunicompo[mult=1,arg=history,anchor1=-160,anchor2=18]{InteractoContext}{UndoRedoHistory}
\umlunicompo[mult=0..*,arg=undos,anchor1=-159,anchor2=150]{UndoRedoHistory}{Undoable}
\umlunicompo[mult=0..*,arg=redos,anchor1=-40,anchor2=25]{UndoRedoHistory}{Undoable}
\end{tikzpicture}
\caption{The metamodel of an \wblit.}\label{fig.abssyntax}
\end{figure}

\emph{An \wblit focuses on transforming the execution of a user interaction into \ui{} command instances}.
The main goal of an \wblit is to produce, update, execute, or cancel one \ui{} command instance along one execution of a user interaction.
\Cref{fig.abssyntax} gives the metamodel of an \wblit.
An \wblit operates within an \wb context that contains an undo/redo history.
Formally, an \wblit $b$ is set between a user interaction $i$ of type $\mathcal{I}$, that exposes an interaction data type $\mathcal{D}$, and a \ui command of type $\mathcal{C}$.
Each interaction execution $exec_i(b)$ of $i$, managed by $b$, may produce a new instance $c$ of the command $\mathcal{C}$, such that:
$exec_i(b) \rightarrow c \land exec'_i(b) \rightarrow c' \implies c \ne c'$.
A user interaction has a behavior, such as an FSM.

\noindent\begin{minipage}{\linewidth}
\begin{lstlisting}[language=MyJava, numbers=left, xleftmargin=2em, caption={The algorithm in pseudo-code of the \wblit behavior. An \wblit operates when its user interaction is running.}, label={lst.algoBinding}, basicstyle={\ttfamily\linespread{1.1}\scriptsize}, emph={when,first,toProduce,end,cancel,then,continuous},emphstyle={\em}]
onInteractionStart() {|\label{code.behav1}|
// when(): states whether the binding can create the command
   if(when()) { |\label{code.behav2}|
      currentCommand = toProduce(); // Creation of the command|\label{code.behav3}|
      first(); // Command initialisation|\label{code.behav4}|
   }
}
onInteractionUpdate() {|\label{code.behav6}|
   if(when()) {
      if(currentCommand == null) {|\label{code.behav7}|
         currentCommand = toProduce();
         first();
      }|\label{code.behav8}|
      then(); // Command update|\label{code.behav9}|
      // The binding executes the command on each update:
      if(continuous() && currentCommand.isExecutable()) {|\label{code.behav19}|
         currentCommand.execute();
      }|\label{code.behav20}|
}}
onInteractionEnd() {|\label{code.behav10}|
   if(when()) {|\label{code.behav11}|
      if(currentCommand == null) {
         currentCommand = toProduce();
         first();
      }
      then();|\label{code.behav12}|
      if(currentCommand.isExecutable()) {|\label{code.behav13}|
         currentCommand.execute(); // Execution of the command
         // Registration of the command (for undo/redo)
         registerCurrentCmd();|\label{code.behav14}|
      }
   }
   end();|\label{code.behav22}|
   currentCommand = null;|\label{code.behav16}|
}
onInteractionCancel() {|\label{code.behav17}|
   cancel();|\label{code.behav23}|
   if(continuous() && currentCommand.wasExecuted()) {
      undoCurrentCommand();
   }
   currentCommand = null;
}|\label{code.behav18}|
\end{lstlisting}
\end{minipage}

\Cref{lst.algoBinding} gives the algorithm of the behavior of an \wblit $exec_i(b)$.
Method calls in italic (namely: \emph{when}, \emph{first}, \emph{toProduce}, \emph{end}, \emph{cancel}, \emph{then}, \emph{continuous}) refer to methods available during the building of the \wblit (\ie with an \wber), as detailed in the next section.
The user interaction life cycle (\Cref{fig.lifeCycleInter}) drives the behavior of an \wblit.
First, when an interaction starts (\Cref{code.behav1}) the \wblit creates a new \ui command if the condition of the \wblit (\emph{when}, \Cref{code.behav2}) is fulfilled.
The \wblit creates a \ui command using the function \emph{toProduce} (\Cref{code.behav3}).
Then, the \wblit initializes the \ui command (\emph{first}, \Cref{code.behav4}).

On each interaction update (\Cref{code.behav6}) the predicate \emph{when} conditions the update of the ongoing \ui command.
One may notice that the \wblit may not create a \ui command when the user interaction starts because of the \emph{when} predicate.
So, if not already created, the \wblit creates a \ui command and initializes it (\Crefrange{code.behav7}{code.behav8}).
The \wblit then updates the \ui command (\emph{then}, \Cref{code.behav9}).
The \wblit executes the ongoing \ui command either at the end of the user interaction, or on each update of the user interaction (what we call \emph{continuous command execution}).
So, if the execution of the ongoing \ui command is \emph{continuous} and the command executable, the \wblit executes it (\Crefrange{code.behav19}{code.behav20}).

When an interaction ends normally (\Cref{code.behav10}), and similarly to the update of the user interaction, the \wblit checks the predicate \emph{when} to possibly create and update the ongoing \ui command (\Crefrange{code.behav11}{code.behav12}).
Then \Cref{code.behav13}, the \wblit checks whether it can execute the \ui command (see the method \emph{isExecutable} \Cref{code.uiCmd2} in \Cref{lst.uiCmd}).
The \wblit puts the executed \ui command in a command register (\Cref{code.behav14}) to keep in memory \ui commands that may be undone by users.
The \wblit finally \emph{ends} (\Cref{code.behav22}) and dereferences the ongoing \ui command (\Cref{code.behav16}).

When the user cancels the ongoing user interaction (\Crefrange{code.behav17}{code.behav18}), the \wblit calls the method \emph{cancel} (\Cref{code.behav23}) and clears the ongoing command (if it exists).

\medskip
\textbf{Start and stop an \wblit.}
An \wblit can start and stop.
On start, an \wblit asks its user interaction to start, \ie to listen for \ui events from the selected interactive objects.
On stop, an \wblit ask its user interaction not to listen for \ui events anymore.
The user interaction also flushes its interaction data.

\subsection{\wber syntax}\label{sub.syntax}

The syntax of the \wber definition language is summarized in \Cref{fig.grammar}.
An \wber works as a builder~\cite{GAM95} to produce one \wblit.
A developer writes an \wber using a set of routines (\emph{config} in \Cref{fig.grammar}) to configure how the produced \wblit will work.
The call to the terminal method \emph{bind} creates and starts the \wblit.
We summarize here the main ones.

\begin{figure}[h]
\small
\centering
\begin{alignat*}{4}
& \textbf{Syntax:}\\ 
& binding & ::= & \textbf{\texttt{binder()}} \overline{config} \texttt{\textbf{.bind()}} & \text{(Binding Configuration)}\\
& config & ::= &\textbf{\texttt{.using(}}() \rightarrow i \textbf{\texttt{)}} & \text{(Interaction Creation)}\\
& & | \quad &\textbf{\texttt{.toProduce(}}d \rightarrow c \textbf{\texttt{)}} &  \text{(Command Creation)}\\
& & | \quad &\textbf{\texttt{.on(}}\overline{w} \textbf{\texttt{)}} &  \text{(Widgets Selection)}\\
& & | \quad &\textbf{\texttt{.first(}}(d, c) \rightarrow void \textbf{\texttt{)}} &  \text{(Interaction Started)} \\
& & |  \quad&\textbf{\texttt{.then(}}(d, c) \rightarrow void \textbf{\texttt{)}} & \text{(Interaction Updated)}\\
& & |  \quad&\textbf{\texttt{.end(}}(d, c) \rightarrow void \textbf{\texttt{)}}&  \text{(Interaction Ended)}\\
& & |  \quad&\textbf{\texttt{.cancel(}}d \rightarrow void \textbf{\texttt{)}} & \text{(Interaction Canceled)}\\
& & |  \quad&\textbf{\texttt{.endOrCancel(}}d \rightarrow void \textbf{\texttt{)}} &  \text{(Interaction Over)}\\
& & |  \quad&\textbf{\texttt{.when(}}d \rightarrow bool \textbf{\texttt{)}} & \text{(Command Condition)}\\
& & |  \quad&\textbf{\texttt{.with(}} \overline{k} \textbf{\texttt{)}} &  \text{(Keyboard Keys)}\\
& & |  \quad&\textbf{\texttt{.throttle(}} int \textbf{\texttt{)}} &  \text{(Throttling)}\\
& & |  \quad&\textbf{\texttt{.strictStart()}} &  \text{(Strict Interaction Start)}\\
& & |  \quad&\textbf{\texttt{.continuous()}} & \text{(Continuous Execution)}\\
& & | \quad &\textbf{\texttt{.consume()}} &\text{(Consume UI Events)}\\
& & | \quad&\textbf{\texttt{.log(}} l \textbf{\texttt{)}} &  \text{(Logging)}\\
& l & ::= & \textbf{\texttt{interaction}} | \textbf{\texttt{binding}} | \textbf{\texttt{cmd}} &  \text{(Logging Level)}\\
& d & & &  \text{(Interaction Data)}\\
& w & & &  \text{(Interactive Object)}\\
& k & & & \text{(Keyboard Code)}\\
& \textbf{Typing:}\\ 
& & & \Gamma \vdash i : \mathcal{I}  & \text{(Interaction Type)}\\
& & & \Gamma \vdash c : \mathcal{C} & \text{(Command Type)}\\
& & & \Gamma \vdash d : \mathcal{D} & \text{(Interaction Data Type)}\\
& & & \Gamma \vdash i :< d &  \text{(Interaction Substitution)}
\end{alignat*}
\caption{The \wber syntax}\label{fig.grammar}
\end{figure}

Using our JavaFX implementation \ijfx, the pseudo-code of \Cref{lst.pseudo} gives the following Java code that describes an \wber that configures an \wblit. 
This \wblit will work using a \emph{DragLock} interaction to produce \emph{Translate} command instances:

\noindent\begin{minipage}{\linewidth}
\begin{lstlisting}[language=MyJava, xleftmargin=2em, caption={An \wber to move a node using a drag-lock}, label={lst.dragLockWB}, numbers=left]
binder()
  .using(DragLock::new)|\label{code.using}|
  .toProduce(d -> new Translate(d.getSrc().getData())) |\label{code.produce}|
  .on(node)|\label{code.on}|
  .first((d, c)-> |\label{code.first}|
      d.getSrcObject().setEffect(new Shadow()))
  .then((d, c) -> c.setCoord(|\label{code.then}|
     c.getShape().getX() + d.getTgtPoint().getX() 
        - d.getSrcPoint().getX(),
     c.getShape().getY() + d.getTgtPoint().getY() 
        - d.getSrcPoint().getY()))
  .when(d -> d.getButton() == MouseButton.PRIMARY)|\label{code.when}|
  .continuous()
  .endOrCancel(d -> d.getSrcObject().setEffect(null))|\label{code.end}|
  .bind();
\end{lstlisting}
\end{minipage}

\smallskip
\noindent\textbf{\emph{using}: User Interaction Creation.}
The routine \emph{using} selects the user interaction the \wblit will use. 
In \Cref{lst.dragLockWB}, \emph{using} takes as argument a function that returns a new instance of the predefined drag-lock interaction (\Cref{code.using}).

\smallskip
\noindent\textbf{\emph{toProduce}: Command Creation.}
The routine \emph{toProduce} (\Cref{code.produce}) focuses on the production of a \ui command.
This routine takes as argument an anonymous function that returns a \ui command as depicted in \Cref{lst.dragLockWB} (\Cref{code.produce}).

\smallskip
\noindent\textbf{\emph{on}: Nodes Selection.}
The user interaction operates on selected interactive objects to produce commands.
The routine \emph{on} allows developers to specify these interactive objects (\eg \Cref{code.on}).

The \emph{on} routine can also take as arguments an observable list of interactive objects $l$.
This allows to dynamically register and unregister interactive objects to/from the \wblit:
when an object $w$ is added to a list $l$, the binding will work for $w$ until $w$ is removed from $l$.
In the following code excerpt, the interaction will operate on each object the canvas will contain, dynamically.

\begin{lstlisting}[language=MyJava]
// on can also take an observable list of nodes
.on(canvas.getChildren()) 
\end{lstlisting}
To support this feature using the standard event processing model, developers would have needed to manually develop the glue code that manages the (un-)registration.

\smallskip
\noindent\textbf{\emph{when}: Command Condition.}
An \wblit constraints the creation, the update, and the execution of a \ui command using the \emph{when} routine (\eg \Cref{code.when}).
The \emph{when} routine is a predicate that takes as argument the current interaction data ($d$) to state whether a \ui command can be created, updated, or executed, as detailed by \Cref{lst.algoBinding}.

\smallskip
\noindent\textbf{\emph{first}, \emph{then}, \emph{end}, \emph{cancel}: Interaction Starts, Updates, Ends, Cancels}.
The \wblit calls the routines:
\emph{first} right after the instantiation of a \ui command;
\emph{then} on each update of the running interaction;
\emph{end} on each normal end of an interaction execution (if the \emph{when} predicate is respected);
\emph{cancel} on each cancellation of the current interaction execution.
They take as arguments the current interaction data ($d$ in the following code) and/or the current \ui command ($c$). \Cref{lst.dragLockWB} illustrates the use of these routines (\Cref{code.first,code.then,code.end}).

\subsection{\wber Properties}\label{sub.props}

An \wber has the following properties:

\smallskip
\noindent\textbf{Type-safe}. The return of each builder routine is typed and constrained by the previously called routines.
For example, the next code selects the user interaction \emph{DragLock}.
So, this user interaction selection constraints the routines used next. 
In this example, the attribute \emph{d} of the routine \emph{when} has the type \emph{FromToData}, which is the interaction data type of the drag-lock interaction.
\begin{lstlisting}[language=MyJava]
binder()
  .using(DragLock::new)
  .when((FromToData d) -> ...)...
\end{lstlisting}
The same reasoning applies to the selected \ui command to produce.
Also, to call the method \emph{bind}, the developer should at least have call the routines \emph{using}, \emph{toProduce}, and \emph{on} beforehand.
   
   This permits to write factory methods that shortcuts the writing of bindings.
For example, using our implementations developers rarely call \texttt{binder().using(DragLock::new)} but instead \texttt{dragLockBinder()} using the factory method that partially builds a binding by selecting the drag lock interaction:
\smallskip

\noindent\begin{minipage}{\linewidth}
\begin{lstlisting}[language=MyJava]
public static PartialBinder<...> dragLockBinder(){
  return new Binder<Interaction<FromToData>, FromToData>()
    .usingInteraction(DragLock::new);
}
\end{lstlisting}
\end{minipage}

\medskip
\noindent\textbf{Immutable}. 
On each routine call, the builder clones itself to return a new builder. 
The goal is to ease builder reuse and factorize \wbers code.
For example, the following code starts by defining a partial binder (\texttt{baseBinder}).
This partial binder is then used to create two binders.
\smallskip

\noindent\begin{minipage}{\linewidth}
\begin{lstlisting}[language=MyJava]
var baseBinder = buttonBinder().on(button).end(...);

baseBinder
  .toProduce(...)
  .when(...)
  .bind();

baseBinder
   .toProduce(...)
   .when(...)
   .bind();
\end{lstlisting}
\end{minipage}

\subsection{Testing \wbslit}\label{sub.test}

The proposed model does not impose testing tools to testers and does not change the way developers write \gls{ui} tests.
Instead, the proposed model comes with specific test oracles and testing facilities that complement classical \gls{ui} testing techniques~\cite{banerjee2013} and oracles~\cite{lelli:hal-01114724}.
We classified our proposed test oracles into two groups:
\wblit test oracles;
\gls{ui} command test oracles.

\medskip
\noindent\textbf{\wblit test oracles.}
The goal of an \wblit is to transform one user interaction execution into a \gls{ui} command.
So, we define a testing oracle: \emph{command produced}.
The \emph{command produced} oracle checks whether a user interaction execution produces a given command type.
\Cref{lst.testwb} depicts this oracle using a \gls{ui} test case that performs a \gls{dnd} to then check whether the expected command has been created.
Because \wbslit are first-class objects, we can observe them in a testing context to collect and analyze the \gls{ui} commands they created.
We built a JUnit5 extension that does this job and provides a \emph{BindingsContext} test parameter (automatically injected) that provides testers with \emph{command produced} assertions.
For example \Cref{code.uitestCmdProd} queries this test parameter to assess that a single command of type \emph{AddShape} was produced during the test execution.

\begin{lstlisting}[language=MyJava, numbers=left, xleftmargin=1.5em, caption={Example of UI test case that uses the \emph{command produced} oracle}, label={lst.testwb}]
@Test 
void dndToAddShape(FxRobot robot, BindingsContext ctx) {
  // ...
  robot.moveTo(...).press(...).moveTo(...).release(...);
  // ...
  ctx.oneCmdProduced(AddShape.class);|\label{code.uitestCmdProd}|
}
\end{lstlisting}

\Cref{lst.testwbKO} is another testing example that checks no command of type \emph{AddShape} are created when executing a specific scenario:

\begin{lstlisting}[language=MyJava, numbers=left, xleftmargin=1.5em, caption={A second example of UI test case that uses the \emph{command produced} oracle}, label={lst.testwbKO}]
@Test 
void dndToAddShapeKO(FxRobot robot, BindingsContext ctx){
  //...
  ctx
    .listAssert()
    .noneSatisfy(cmd -> cmd.ofType(AddShape.class));
}
\end{lstlisting}

\medskip
\noindent\textbf{\gls{ui} command test oracles.}
Reifying \gls{ui} commands as first-class objects permits the design of dedicated testing facilities.
First, we define \gls{ui} command oracles implemented in a dedicated testing framework.
Second, we provide testers with a \gls{ui} command test skeletons generator that help them in starting their \ui command testing tasks.

\smallskip

\emph{\gls{ui} command testing framework.}
We defined the following \emph{\gls{ui} command test oracles}:
\emph{can do}, \emph{cannot do}, \emph{do}, and \emph{undo}.
The \emph{can do} oracle checks that the command can be executed.
The \emph{cannot do} oracle checks scenarios where the command cannot be executed.
The \emph{do} oracle checks the correct (re-)execution of the command.
The \emph{undo} oracle checks that the undoable command was correctly undone after its execution.
Note that the \emph{redo} oracle is the same than the \emph{do} oracle.

A dedicated testing framework has to alleviate the job of developers by automating the execution of \gls{ui} commands to help these developers in focusing on writing \gls{ui} commands fixtures and oracles. 
The code of \Cref{lst.testCmd} is an example of a \gls{ui} command test class.
Testers have to write \emph{do} and \emph{undo} test oracles (\Crefrange{code.uitest1}{code.uitest2}). 
Testers must also write \gls{ui} command fixtures for configuring a command that:
can be executed (used for the \emph{can do}, \emph{do}, \emph{undo} oracles), \Crefrange{code.uitest3}{code.uitest4};
cannot be executed (for the \emph{cannot do} oracle), \Crefrange{code.uitest5}{code.uitest6}.
As several \emph{can do} and \emph{cannot do} scenarios may exist, the methods return a set of (\texttt{Stream.of()}) oracles and fixtures.

 \noindent\begin{minipage}{\linewidth}
\begin{lstlisting}[language=MyJava, xleftmargin=2em, numbers=left, caption={Example of a UI command test class}, label={lst.testCmd}]
class DelShapesTest extends UndoableCmdTest<DelShapes> {
  List<Shape> shapes;
  Drawing drawing;
  @Override protected Stream<Runnable> canDoFixtures() {|\label{code.uitest3}|
    return Stream.of(() -> {
      shapes = List.of(Factory.newRec(),
          Factory.newRec(), Factory.newRec());
      drawing = Factory.newDrawing(shapes);
      cmd = new DelShapes(drawing, 
          List.of(shapes.get(0), shapes.get(2)));
    });
  }|\label{code.uitest4}|
  @Override
  protected Stream<Runnable> cannotDoFixtures(){|\label{code.uitest5}|
    return Stream.of(() -> {
      cmd=new DelShapes(Factory.newDrawing(), List.of());|\label{code.uitest6}|
    });
  }
  @Override protected Stream<Runnable> doCheckers() {|\label{code.uitest1}|
    return Stream.of(() -> {
      assertThat(drawing.size()).isEqualTo(1);
      assertThat(drawing.getShapeAt(0))
          .isSameAs(shapes.get(1));
    });
  }
  @Override protected Stream<Runnable> undoCheckers() {
    return Stream.of(() -> {
      assertThat(drawing.size()).isEqualTo(3);
      assertThat(drawing.getShapes()).isEqualTo(shapes);
    });
}}|\label{code.uitest2}|
\end{lstlisting}
 \end{minipage}

\begin{figure}[h]
   \centering
   \includegraphics[scale=0.25]{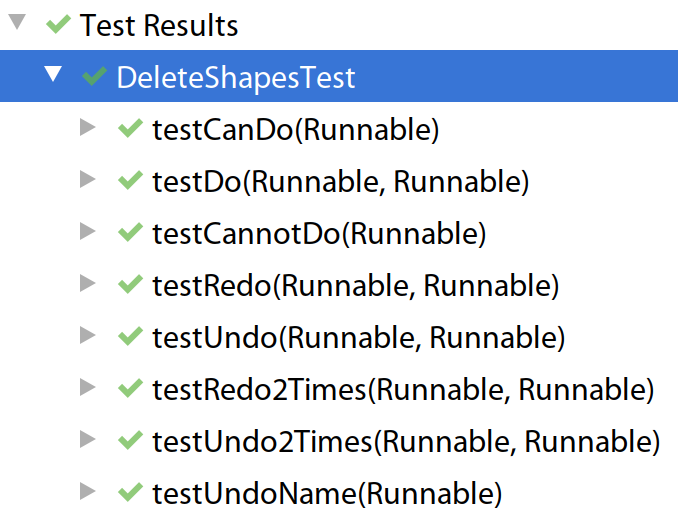}
   \caption{The tests executed by the UI command testing framework for the test class of \Cref{lst.testCmd}}
   \label{fig.uicmdtestres}
\end{figure}

The execution of this test class runs the targeted \gls{ui} command under several scenarios (test instances) to check the \gls{ui} command oracles.
\Cref{fig.uicmdtestres} shows the different tested scenarios: can execute, cannot execute, execute, undo, redo, several do/undo/redo sequences.

\emph{\gls{ui} command test skeletons generator.}
Since \gls{ui} commands are first-class objects, a static analysis can extract from them information to generate \gls{ui} command test class skeletons.
For example with \Cref{lst.testCmd}, testers only wrote the code inside the \texttt{Stream.of()} instructions, the rest being automatically generated.
The test class attributes are copied from the \gls{ui} command class attributes.
The static analysis checks whether the \gls{ui} command is undoable for generating methods \emph{undoCheckers}.

\section{Evaluation}\label{sec.eval}

In this section we evaluate five aspects of \wb:
Is \wb implementable on different \ui platforms?
Does the approach scale in terms of performance and expressiveness?
What are the pros and cons based on our usage?
What is the scope of \wb?
Can students successfully use \wb?

All the material of this section is available on our companion web page\footnote{\url{https://github.com/interacto/research}}.

\subsection{Implementations}\label{eval.implem}

We define the first research question we address as follows:

\smallskip

\noindent \textbf{RQ1.}
Can we implement the proposed model in different programming languages and graphical toolkits that support different paradigms?

\medskip
We implemented the \wb approach on the top of two programming languages:
Java and TypeScript~\cite{bierman14}.
For each of these two languages we provide a \ui platform-independent library, namely \ijava and \its (this last uses the native Web graphical API\footnote{\url{https://developer.mozilla.org/en-US/docs/Web/API}}).

Regarding Java, we implemented an extension library, \ijfx, for supporting the JavaFX \ui toolkit~\cite{jfx}, a major \ui toolkit for Java.
\emph{Interacto-JavaFX-Test} complements \ijfx with testing facilities for JavaFX.
The support of another Java UI toolkit, such as Android, would require an new library that extends \ijava by defining how to register to low-level \ui events of the \ui platform.

Regarding TypeScript, we implemented an extension library, \iang, for improving the use of \its within the \emph{Angular} UI toolkit~\cite{angular}.

The implementations contain around \SI{8000}{lines} of Java~11 code and \SI{5000}{lines} of TypeScript~3.8 code.
Both implementations rely on the same concepts detailed in this paper.
User interactions are developed using composite and concurrent \glspl{fsm}. 
As illustrated in \Cref{sec.motiv} with \Cref{fig.dragLockAlt}, the drag-lock FSM is composite:
its transitions labelled \emph{double click} refers to the double-click interaction.
Regarding concurrent FSMs, an example is the multi-touch interaction where each touch that starts corresponds to touch interaction\footnote{See the companion web page for more details on the multi-touch interaction.}.
Both implementations provide a set of around \num{30} predefined user interactions that developers can use.
The implementations are open-source and free available\footnote{\url{https://interacto.github.io}}.

We implemented several optimizations in the implementations of user interactions for both Java and TypeScript:
\begin{itemize}
   \item \textbf{Efficient event registration}.
      For example with \Cref{fig.dragLockAlt}, the \gls{fsm} on the top uses \emph{mouse click} (for double click), \emph{mouse move}, and \emph{key pressure} events.
   When a user interaction is activated, it does not listen for all the possible \gls{ui} events that this user interaction uses.
   Instead, when entering a state the accepted \gls{ui} events at this state are identified, and the user interaction registers for these \gls{ui} events only (and un-register for the other ones).
   For example, when the top FSM of \Cref{fig.dragLockAlt} starts, it only listens for mouse click events (as the fist event of a double click interaction).
   When entering the state \emph{Locked}, this FSM now listens for mouse move and mouse click events only.
   When programming user interactions by hand, software engineers have to think about this optimization to then manually craft it.
   \item \textbf{Late starting}.
   Let us take the example of the \gls{dnd}.
   One developer may consider that a \gls{dnd} starts at the first move, not at the initial mouse pressure.
   In such a case, one can specify the state of a user interaction that will correspond to the starting of the user interaction (that calls the \emph{map} and \emph{first} routines).
   \item \textbf{\wber shortcuts}.
   Our implementations provide shortcuts for initializing \wbers.
   For example, \Cref{lst.tsExample} contains two \wbers coded in TypeScript with the Angular framework (to be discussed later in this section).
   The first \wber starts with \texttt{multiTouchBinder(2)}.
   This code is equivalent to \texttt{binder().using(() => new MultiTouch(2))}:
   for most of the user interactions we defined, we provide a coding shortcut for initializing \wbers that use a given user interaction.
   In this case, the user interaction multi-touch is configured to use two touch points.
\end{itemize}

\noindent \textbf{Differences between the Java and TypeScript implementations.}
Regarding user interactions, each graphical toolkit has its own \gls{ui} events.
If most of such \gls{ui} events are common across the graphical toolkits, some others are platform-specific or have different parameters.
For examples:
the touch event of the native Web API has a force parameter (the amount of pressure the user is applying) that is not supported in JavaFX;
JavaFX provides window-based \gls{ui} events that the native Web API does not provide;
the native Web graphical toolkit has a multi-touch support so that our TypeScript implementation provides various standard touch user interactions such as \emph{pan}, \emph{swipe}, \emph{tap}.
This has a limited impact of the library of predefined user interactions provided by each implementation.

Still related to user interactions, the Angular/Web API has a feature that permits to disable the default user interaction initially imposed by the Web browser.
For example, a right-click shows a menu on most of Web browser, which can enter conflict with the expected behavior of a developed Web application.
This feature is called \emph{preventDefault}.
We support this feature for user interactions of our TypeScript implementation.
For example with \Cref{lst.tsExample}, we used this feature for the first \wber \emph{multiTouchBinder}.

The last difference concerns the name of specific \wber routines.
The name of several concepts differ from JavaFX to Angular/Web API.
For example, the concept that consists of stopping the propagation of a UI event refers to the term \emph{consume} on JavaFX and to the term \emph{stopImmediatePropagation} on Angular/Web API.
Our implementations use the name used by their UI toolkit.
Similarly, because of TypeScript method conflicts, the TypeScript implementation uses the name \emph{onDynamic} (see \Cref{lst.tsExample}) to refer to the \emph{on} routine that takes as argument a list to observe.
\medskip

\noindent\textbf{Example of how \ijfx works within a JavaFX controller and an Angular component.}

\wb can work with any architectural pattern that processes \gls{ui} events and provides access to the user interface elements in the code.
We illustrate this point with the two following Angular and JavaFX examples.

\smallskip
\noindent\begin{minipage}{\linewidth}
\begin{lstlisting}[language=MyTS, numbers=left, xleftmargin=2em, caption={An example of how developers can code \wbers within an Angular component}, label={lst.tsExample}]
export class AppComponent implements AfterViewInit {
	@ViewChild('canvas') private canvas: ElementRef;~\label{code.ts1}~
	// ...
	ngAfterViewInit(): void {~\label{code.ts2}~
		multiTouchBinder(2)~\label{code.ts3}~
			.toProduce(i => new DrawRect(
				this.canvas.nativeElement as SVGSVGElement))
			.on(this.canvas.nativeElement)
			.then((c, i) => {
				c.setCoords(Math.min(...i.getTouchData()
				  .map(touch => touch.getTgtClientX()))-b.x,...);
			})
			.continuousExecution()
			.preventDefault()
			.bind();
		
		tapBinder(3)~\label{code.ts4}~
			.toProduce(i => new ChangeColor(
				i.getTapData()[0].getSrcObject()))
			.onDynamic(this.canvas.nativeElement)
			.when(i => i.getTapData()[0].getSrcObject() 
				!== this.canvas.nativeElement && ...)
			.bind();
}}
\end{lstlisting}
\end{minipage}

\Cref{lst.tsExample} is an excerpt of an Angular component (a kind of controller in the Angular framework).
It shows how one can use \iang within an Angular component.
Currently, \iang requires the widgets to be accessible from the component code.
This requires the definition of a class attribute that corresponds to a widget defined in the HTML part of the component (the view of an Angular component is an HTML document).
To overcome this requirement, we can improve \iang with dedicated Angular facilities.
In this example, \emph{canvas} refers to such a widget (\Cref{code.ts1}).
\wbers are defined in the method \emph{ngAfterViewInit} (\Cref{code.ts2}), which is a special method of each Angular component:
Angular automatically calls this method once the HTML view of the component loaded, so that \wbers can access its widgets.
This method contains two \wbers:
the first one (\Cref{code.ts3}) builds an \wblit that uses a multi-touch interaction to produce \emph{DrawRect} commands;
the second one (\Cref{code.ts4}) builds an \wblit that uses a tap interaction to produce \emph{ChangeColor} commands.
These commands are specific to this application and defined in a dedicated folder.
The code of these \wbers and their the commands they use is similar to the Java code show in this paper to detail and illustrate the proposal.

\medskip

\Cref{lst.jfxcontrol} gives a concrete example about how \wbers are defined in a JavaFX controller.
The \emph{CanvasController} class is in charge of managing a JavaFX view (not depicted here).
The JavaFX dependency injection permits the controller to access the interactive objects of its view (\cf the annotation \emph{@FXML}, \Cref{code.jfx1}).
JavaFX calls the method \emph{initialize} at the first use of the controller.
We use this method (\Cref{code.jfx2}) to code two \wbers that configure JavaFX \wbslit (\Cref{code.jfx3,code.jfx4}).

\smallskip
\noindent\begin{minipage}{\linewidth}
\begin{lstlisting}[language=MyJava,numbers=left, xleftmargin=2em, caption={Example of how developers can code an \wber within a JavaFX controller}, label={lst.jfxcontrol}]
public class CanvasController ... {
	@FXML Canvas canvas;|\label{code.jfx1}|
	// ...
	public void initialize(...) {|\label{code.jfx2}|
		binder()|\label{code.jfx3}|
			.using(DragLock::new)
			...
			.on(canvas.getChildren())
			.bind();
		binder()|\label{code.jfx4}|
			.using(DnD::new)
			.toProduce(d -> new Add(...))
			.on(canvas).
			...
			.bind();
}}
\end{lstlisting}
\end{minipage}

\medskip

\textbf{To conclude on RQ1, these two implementations detailed in this section show that the proposed model is not tied to one specific programming language or graphical framework.}

\subsection{A real world use case: \SOFTWARE}\label{eval.case}

\noindent \textbf{RQ2. Does the implementation of the proposed model scale?}
We first discuss the ability of the proposal to scale for the development of a representative software system.
Then, we discuss performance of the implementation compared to the use of the standard \gls{ui} event processing model. 

\medskip

\SOFTWARE is a large open-source and highly interactive vector drawing editor for \LaTeX{}\footnote{\url{http://latexdraw.sourceforge.net}}.
It is downloaded \dltimes times per month.
\SOFTWARE is distributed on more than \num{10} Linux distributions, also available on Windows and MacOS.
On its Github page, the project has \num{340} stars and \num{58} forks.
\SOFTWARE is composed of around \num{35000} lines of Java code.
We developed \SOFTWARE in Java since 2005 using callback methods for processing \gls{ui} events.
As any software system, \SOFTWARE evolves:
we progressively introduced the use of \ijfx instead of callback methods since 2013.
The current version of \SOFTWARE (4.0) now entirely relies on the fluent \wber API.
Note that the development of \SOFTWARE precedes the development of \ijfx (2005 \emph{vs} 2013).
Moreover, the goal of \SOFTWARE has no relation with \wb.
Because of the successive evolutions, the fully callback version and the fully \wb version are not isomorphic and cannot be compared.
\SOFTWARE~4.0 has the following characteristics.
It is composed of \num{45} controllers (around \num{4700} lines of Java code) that contain a total of \num{224} \wbslit (around \num{1000} lines of Java code).
These \wbslit produce \num{37} different \gls{ui} commands (around \num{1400} lines of Java code).
Regarding the used user interactions:
\num{45} \wbslit use several forms of \gls{dnd} (\num{12}), click (\num{11}), or keyboard interactions (\num{22}).
The rest of the \wbslit (\num{179}) are based on standard widgets and window interactions (buttons, lists, color pickers, \etc).
We developed all the \wbslit using the \wber fluent API described in this paper.
On the issue tracking system of the Github page of the project, eight issues, on the \num{30} ones opened, were related to user interaction processing.

\medskip
Regarding the testing oracles and features we propose with \ijfx and \emph{Interacto-JavaFX-Test}, we tested all the \ui command of \SOFTWARE using the proposed \ui command oracles.
We used the \ui command test generator to produce test class skeletons we then completed.
Before using this testing feature we wrote only few \ui command test classes.
The generated test classes replaced these test classes as we considered the new ones as covering more UI command concepts (undo, redo, \etc) and they achieve a better code coverage (currently \SI{98.3}{\percent} of covered lines).
We, however, cannot currently use the \wblit oracle for testing \SOFTWARE's \wbslit as this testing feature requires JUnit5 tests while \SOFTWARE UI test suite is written in JUnit4.
We use this testing feature for testing the library \ijfx and in other cases as discussed in the next section.

\smallskip
\textbf{We use \ijfx since 2013 for developing a highly interactive and widely-used open-source software system.
This shows the ability of the proposal to scale for the development of such software systems.}

\medskip

\begin{figure}[h]
   \centering
   \includegraphics[width=0.8\columnwidth]{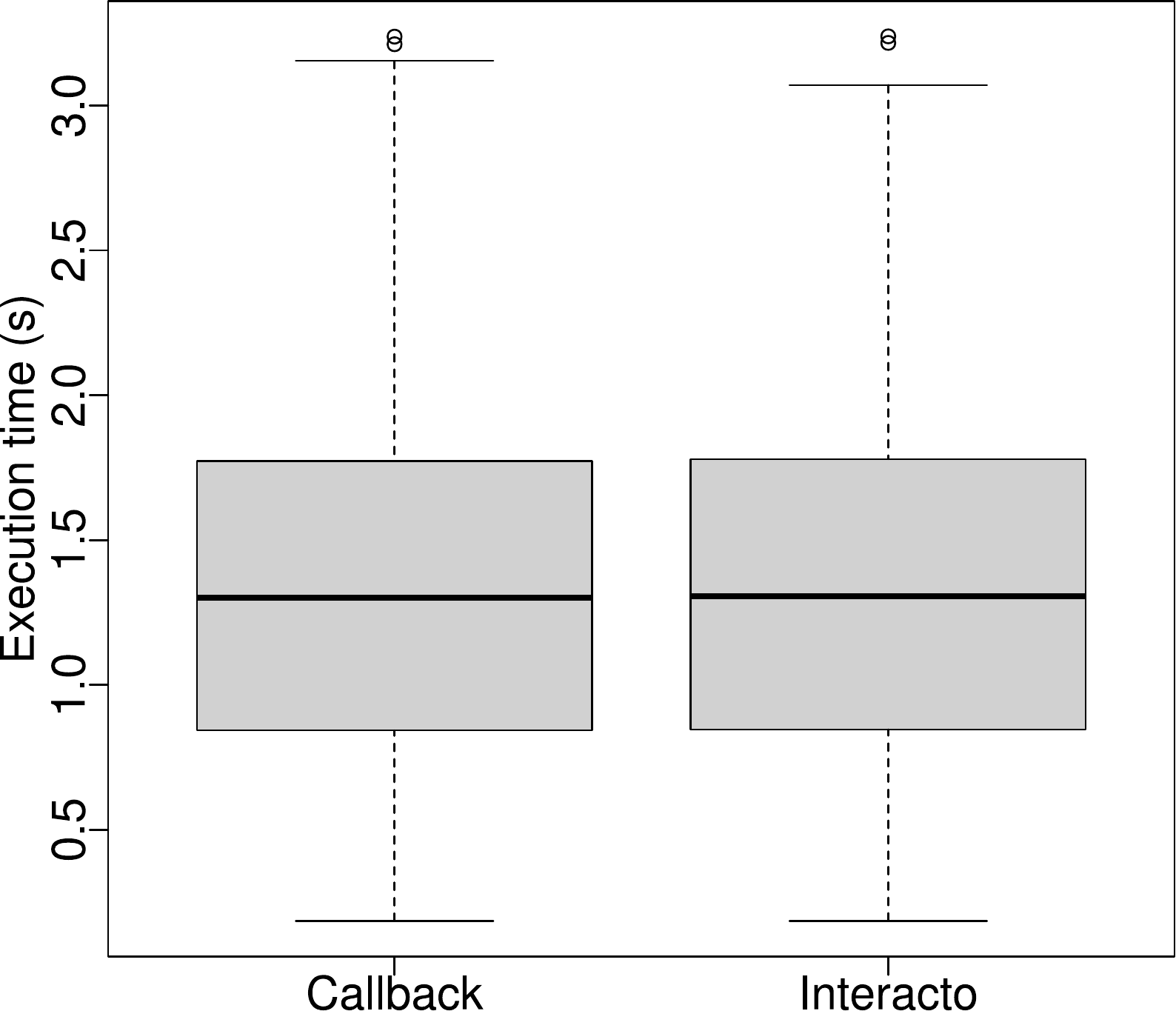}
   \caption{Execution time comparison}\label{fig.perfs}
\end{figure}

Regarding the performance of the implementation, we evaluated the possible overhead of the use of \ijfx compared to the use of \gls{ui} callbacks.
Because of the successive evolutions of \SOFTWARE, the fully callback version and the fully \ijfx version are not isomorphic and cannot be compared.
We thus ported \SOFTWARE~4.0 to use \gls{ui} callback methods only.
This new callback version still uses the developed \gls{ui} commands.
The controllers of this callback version have the following code metrics:
\num{5600} lines of code (vs \num{4700} lines for the \ijfx version);
A cyclomatic complexity (mean per method) of \num{2.55} (vs \num{1.89});
An LCOM value (Lack of Cohesion
in Methods) of \num{1.86} (vs \num{1.49}).
We used the existing test suite of \SOFTWARE to validate the call back version:
the \SOFTWARE test suite, that covers \SI{90}{\percent} of the \gls{ui} controllers code and \SI{100}{\percent} of the \wbers code, assesses the isomorphic property between the \ijfx and callback versions.
Then, using these two implementations (\ijfx and \gls{ui} callback), we executed \SI{10}{times} the test suite that covers the \wbers.
We used a Linux computer with \SI{4}{CPUs} of \SI{2.6}{\GHz} each, \SI{16}{\giga\byte} RAM, an Intel HD Graphics 5500 graphical card, and Xorg~1.19.6 as graphical server (the used test suite is composed of user interface tests), and Java~8 update~161.
The test execution was not parallelized.
We removed all the tests not related to \gls{ui} event processing.
We obtained a test suite composed of \num{447} tests.
These GUI tests simulate user interactions with the \gls{gui} to trigger (or not) the creation of commands.
So these tests directly operate on \wbers and UI callbacks.
We modified the resulting test suite to automatically measure and log the execution time of each test.
We removed the assertions of these tests to transform them into executable \gls{ui} scenarios only.
\Cref{fig.perfs} summarized the measured execution times.

For each test we computed its mean execution value based on ten executions.
The mean execution time for a test is \SI{1.331}{\s} for the callback version and \SI{1.334}{\s} for the \ijfx version (see \Cref{fig.perfs}).
We use the Wilcoxon signed-rank test (a paired difference test)~\cite{Sheskin2007} to compare the two sets of mean execution times (data do not follow a normal distribution and we used an initial confidence level of \SI{95}{\percent}, \ie $\alpha = 0.05$):
the observed differences between the callback and \ijfx versions (\emph{p}-value of \num{0.9472}) are not significant and may be due to randomness.
We can conclude that using the JavaFX implementation the conducted experiment gives no significant evidence regarding an execution time overhead caused by the use of \ijfx compared to the implementations based on callback methods.

\textbf{This experiment shows no overhead in terms of performance when using \ijfx compared to standard JavaFX \gls{ui} callbacks.}

\bigskip

\noindent \textbf{RQ3. What are the pros and cons of the proposed model based on our real world usage?}
This research question permits to discuss about the scope, the expressiveness of our proposal mainly in the context of \SOFTWARE.

The use of \ijfx in \SOFTWARE completely removed the use of \gls{ui} callbacks for processing \gls{ui} events.
We faced no situation in which we could not use \ijfx to process all the heterogeneous user interactions that \SOFTWARE employs.

The use of \ijfx does not remove the use of general-purpose callbacks and data binding features in \SOFTWARE.
The first reason is that the \wb approach complements data binding and achieves different purposes:
\wb provides features for processing user interactions while data binding creates dynamic links between object values.
For example, the following code statement, from \SOFTWARE, establishes a data binding that does not imply the use of any \gls{ui} command or user interaction:

\smallskip
\noindent\begin{minipage}{\linewidth}
\begin{lstlisting}[language=MyJava2]
msg.visibleProperty().bind(Bindings
  .createBooleanBinding(!msg.getText().isEmpty() ...));
\end{lstlisting}
\end{minipage}

Similarly, the following reactive programming code, extracted from \SOFTWARE, listens for changes in an observable list to call the method \emph{updateSelectionBorders}.
This reactive code uses throttling to limit the number of updates and improve performance.
This code does not refer to any \gls{ui} event (and thus not to any user interaction) and is thus out the scope of \wb.

\smallskip
\noindent\begin{minipage}{\linewidth}
\begin{lstlisting}[language=MyJava2]
JavaFxObservable.<ObservableList<Shape>>changesOf(
        drawing.getSelection().getShapes())
  .throttleLast(20, TimeUnit.MILLISECONDS)
  .observeOn(JavaFxScheduler.platform())
  .subscribe(next -> updateSelectionBorders());
\end{lstlisting}
\end{minipage}

These two remarks also concerns the Angular framework that supports data binding and reactive programming features.
\smallskip

The main drawback we found concerns several complex \wbers that require temporary variables to share data across the different routines of the binding.
For example, the following \wber code defines and uses two objects, namely \emph{xgap} and \emph{ygap}, to share data across the \emph{first} and \emph{then} routines.
In this case, developers may prefer to define a new class instead of using an \wber.
This is possible with our implementations.

\smallskip
\noindent\begin{minipage}{\linewidth}
\begin{lstlisting}[language=MyJava]
private void configureDnD2ScaleBinding() {
	final AtomicInteger xgap = new AtomicInteger();
	final AtomicInteger ygap = new AtomicInteger();
	binder()
		...
		.first((i, c) -> {
			// This routine sets values to xgap and ygap.
		})
		.then((i, c) -> {
			// This routine uses xgap and ygap.
		})
		.bind();
}
\end{lstlisting}
\end{minipage}

The code above also shows that we placed the \wber code into a specific method.
Several controllers have multiple and complex \wbers.
Putting all the \wbers in the same method would produce a long method code smell.
For readability we split these definitions into different methods with a name that clearly describes the job of the \wber.

\medskip
Another point is related to coding style.
There exists two ways for registering callbacks on widgets in both JavaFX and Angular:

\noindent 1/ \textbf{Programmatically}, one can register as follows in this TypeScript code:

\begin{lstlisting}[language=MyTS]
button.addEventListener('mousedown', evt => {...});
\end{lstlisting}

\noindent 2/ One can also register \textbf{in the UI description code}, here in the HTML code of an Angular application:

\begin{lstlisting}[language=MyHTML]
<button (mousedown)="myCallback()"/>
\end{lstlisting}
where \emph{myCallback} is a TypeScript method defined in the code. The goal of this style is to permit stakeholders with limited programming skills (\eg designers) to specify which user interactions to use into views.

The \wb implementation for Angular supports both styles.
For example, the following Angular HTML example defines an SVG rectangle on which the use of a drag-lock is specified for moving it (Interacto attribute \emph{ioDraglock}):

\begin{lstlisting}[language=MyHTML]
<rect [ioDraglock]="moveRect"></rect>
\end{lstlisting}

The method \emph{moveRect} is defined in the Angular component as follows.
This method takes as arguments a partial binder: the user interaction to use is already configured.

\begin{lstlisting}[language=MyTS]
moveRect(binder: PartialDragLockBinder): void {
  binder
    .toProduce(() => ...)
	....bind();
}
\end{lstlisting}

\medskip
Finally, in some specific cases a developer may not want to rely on \gls{ui} events but on data binding as an alternative to \gls{ui} event processing.
For example, the following code establishes a binding between the text value of a text field, changed when a user interacts with the text field.
\begin{lstlisting}[language=MyJava2]
msg.textProperty().bind(button.textProperty());
\end{lstlisting}
Instead of processing \gls{ui} keyboard events produced by this text field, a developer can bind the text value to another value.
This way of binding data, however, does not support undo/redo features for JavaFX and Angular.
Conceptually and technically, \wb can support such a use case. 
To do so, we have to build a user interaction whose FSM uses the value change event of such text properties. 
Then, we can use this new interaction within an \wblit. 
However, the core idea of \wb to focus on real user interactions.

\subsection{Scope of the proposed model}\label{eval.scope}

We discuss here the expressiveness and extensibility of the proposal through the kinds of software systems it can support.

\medskip
\noindent \textbf{RQ4.
What is the scope of the proposed model?}

This research question aims at discussing the types of \glspl{ui} \wb supports and to what extent it is extensible.
\medskip

The long-term use case detailed in \Cref{eval.case} discusses about \SOFTWARE, a highly-interactive graphical software systems.
\SOFTWARE employs both a large panel of standard widgets (\eg buttons, checkboxes, lists, tabs) and 2D user interactions dedicated to the handling of 2D shapes.
\SOFTWARE must be used with mice and keyboards.

We use \ijfx to build other yet smaller software systems.
We developed and maintain \spoonVis, a graphical tool for visualizing and interacting with the Spoon \gls{ast} of Java code.
Spoon is a Java framework for developing dedicated Java code analyzers~\cite{pawlak:hal-01169705}.
It parses Java code to build a Spoon \gls{ast} that one can handle to transform or analyze Java code.
\spoonVis uses Spoon to parse Java code and display the resulting Spoon \gls{ast} using tree-based and text widgets.
\spoonVis uses other user interactions and widgets that \SOFTWARE and covers a different domain.
The Spoon maintainers accepted \spoonVis and merged it in the main branch of Spoon as a tool of the Spoon ecosystem\footnote{\url{https://github.com/INRIA/spoon/tree/master/spoon-visualisation}}.

Regarding \iang, we ported \spoonVis to Angular\footnote{\url{https://github.com/arnobl/spoon-web/}}.
We also develop an illustrative Angular Web application to explain how to use the TypeScript implementation of our proposal within Angular\footnote{\url{https://github.com/interacto/example-angular}}.
This application mainly uses touch-based user interactions (multi-touch, swipe, \etc).

These developments show that \wb can be used in other contexts that \SOFTWARE.
The types of software systems one can develop using \wb are the same as the ones targeted by the graphical toolkits we support, namely Angular and JavaFX.
So, this encompasses desktop, Web, mobile software systems developed to use touch-screens, mice, keyboards, as input devices.
We do not see any blocking issue for porting our proposal to other graphical toolkits that operates with object-oriented programming languages such as Android or React as they rely on the same concepts as Angular and JavaFX.

Regarding the extensibility of the proposal, UI developers can develop new user interactions for \wb implementations using the atomic \gls{ui} events provided by the \gls{ui} toolkit.
To do so, the developer has to create one class for the user interaction and a second class for its FSM.
The developer may have to create another class for defining the data the new user interaction will expose (if the existing interaction data classes do not match the requirements).
If a developer has to support a new UI events, he has to create a new class that represents a transition to be used in FSMs.
The process is the same for Angular and JavaFX.
The companion web page points to the code of a very simple example.

We think that these steps may take more time for a developer than coding the classical assembly of \gls{ui} events.

Regarding the integration of \wb within guidelines of UI toolkits, our implementations aim at following the same programming style that the one proposed by the UI toolkits. For example with Angular, \wb works with pre-configured Angular linters, leverages Angular standard features (\eg dependency injection, services, directives), and regarding event processing permits both coding styles.
Moreover, style guides of UI toolkits mainly focus on design choices to follow. 
For example, double-click interactions are not allowed in Gnome applications\footnote{\url{https://developer.gnome.org/hig/stable/}}.
These guides have no impact on \wb since they do not discuss the way user interactions are technically processed.

\subsection{An empirical study with students}\label{eval.std}

Five contributors developed the long-term use case we detailed in \Cref{eval.case}.
To overcome this threat of generalization we now discuss the following research question:

\smallskip
\noindent \textbf{RQ5. To what extent beginners successfully use \wb for processing \gls{ui} events compared to standard \gls{ui} toolkits?}
\smallskip

To study this research question we conducted an empirical study that involved students.

\medskip
\noindent \textbf{Objects.}
The object of the experiments is an Angular~9.1 Web application we created for the experiment.
This application has a simple Angular service that stores data.
This application also has a single Angular component that uses this service.
The HTML document of this component contains several widgets:
undo/redo buttons; a \emph{div} tag that contains text ; a text area; an SVG document that contains one rectangle.
The Angular component does not contain any code related to the processing of \gls{ui} events produced by these widgets.

We duplicated the application to have two applications for the experiment:
a classical Angular application;
the same Angular application for which we added \iang~5.3.0 in its dependencies.

We focus on Angular (and the underlying native Web API) and Interacto for two reasons:
\iang is implemented in TypeScript and works with Angular;
Angular is a major Web application framework widely used in the industry.
It relies on major software engineering concepts, in particular reactive programming through its data binding features for overcoming the limits of the \emph{Observer} pattern.

\bigskip

\noindent \textbf{Subjects.}
The subjects of the study are 44 master students in computer science with a strong focus on software engineering.
Studies~\cite{salman2015,svahnberg2008} showed that students can be valid and well representative subjects for experiments and development tasks.
The subjects are all volunteers with a background on Web development with Angular.
Each subject answered three questions regarding his/her expertise in: programming in general (q1); Web programming in general (q2); Web programming with Angular (q3).
They have to give a number between 1 and 10 included (1 meaning no expertise and 10 very strong expert).
We used their answers to form two balanced groups of 22 students each:
group G1 that performed the experiment with Interacto;
group G2 that used Angular.
To form G1 and G2 we paired subjects that provided similar answers to the three questions to obtain 22 pairs of subjects. 
G1 and G2 each contain one subject of each pair to balance the two groups.
\Cref{tab.groups} details the mean values of the groups for the three questions.

\begin{table}[h]\small
	\centering 
	\caption{Expertise of each group of subjects (mean values)}\label{tab.groups}
	\begin{tabular}{cccc}
		\toprule
		\textbf{Group}& \textbf{q1} & \textbf{q2} & \textbf{q3}\\
		\cmidrule(lr){1-1}\cmidrule(lr){2-2}\cmidrule(lr){3-3}\cmidrule(lr){4-4}
		G1 & \num{6} & \num{3.8} & \num{4.1}\\
		\cmidrule(lr){1-1}\cmidrule(lr){2-2}\cmidrule(lr){3-3}\cmidrule(lr){4-4}
		G2 & \num{5.8} & \num{3.7} & \num{3.8}\\
		\bottomrule
	\end{tabular}
\end{table}


\noindent \textbf{Sub Research Questions.}
To discuss RQ5, we formulate the three following sub research questions:

\noindent \textbf{RQ5.1} Does the use of Interacto improve the correctness to process \gls{ui} events on typical development tasks?
\smallskip

\noindent \textbf{RQ5.2} Does the use of Interacto reduce the time on those tasks?
\smallskip

\noindent \textbf{RQ5.3} To what extent the students prefer using Interacto for completing those tasks?

\bigskip

\noindent \textbf{Tasks.}
We designed three representative tasks that cover different UI development aspects, in particular user interaction usages and undo/redo support.
The three tasks use different user interactions more or less complex to use or code.
Each group had to do the same three tasks (namely, T1, T2, T3).
We established a time limit for each task:
if reached, they have to stop working on their current task and commit the changes.
The total duration of the session was \SI{95}{\minute}.

\emph{T1.}
This task is composed of two sub-tasks~T1.1 and~T1.2.
The goal of T1.1 is to use a simple user interaction, the triple click, that subjects can easily use in Angular and Interacto.
The subject had to use a triple-click on a given HTML \emph{div} tag to change its color (stored in an Angular service).
The goal of T1.2 is to support the undo/redo of this color change.
The time limit of T1 is \SI{35}{\minute}.

\emph{T2.}
This tasks focuses on the use of a more complex user interaction.
This interaction consists in typing text in a text area.
If the user stops writing after a delay of 1 second, the text data stored in the Angular service must be updated.
This is a mainstream user interaction that text processing tools use to limit the number of editing actions.
The time limit of T2 is \SI{20}{\minute}.

\emph{T3.}
This task is composed of two sub-tasks~T3.1 and~T3.2.
The goal of T3.1 is to use a complex user interaction, the DnD, provided by Angular and Interacto.
The subjects had to use a DnD to move a rectangle.
The Angular service stores the rectangle data.
A 2D SVG rectangle renders graphically these data.
Angular provides DnD features for moving objects graphically by changing, for example, its graphical style (its CSS).
This Angular feature, however, does not modify the possible data model a dragged object renders, here the rectangle data.

Interacto provides a DnD interaction, but subjects had to employ it to change the coordinates.
The goal of T3.2 is to support the undo/redo of this move.
The time limit of T1 is \SI{40}{\minute}.

In this section we refer to \emph{T-UNDO} as the tasks related to undo/redo operations, namely: T1.2 and T3.2.
We also refer to T-UI as the tasks related to user interactions, namely: T1.1, T2, and T3.1.
The goal of these transversal tasks is to discuss results by topics rather than tasks.

\bigskip

\noindent \textbf{Dependent Variables.}
We collected or computed the following variables:
\begin{itemize}
	\item \emph{Average Time} (TIME): 
	measures the average time in minutes the subjects spent to complete each question.
	We computed this metric based on the time stamps of the commits each subject made:
	we asked the subjects to commit locally before and after each question.
	
	\item Correct Answer (CORR): 
	measures the correctness of each question answered by a subject.
	We measured CORR by designing 10 \gls{ui} tests:
	four tests for T1 and T3 that test the user interaction, the data changes, the undo, and the redo;
	two tests for T2 that test the user interaction and the data changes.
	The result of a test execution is a boolean value.

	\item Level of Difficulty (DIFF): 
	measures the difficulty felt by each subject for each task.
	After each task (or sub-task), the subject gave the associated DIFF between 1 and 10 included	where 1 is a very easy task and 10 a very difficult one.
\end{itemize}

\bigskip

\noindent \textbf{Experimental Protocol.}
The subjects had no knowledge about Interacto before the experiment.
They followed several courses of Web development using Angular and may have skills on Angular acquired during internships in the industry.
To reduce this knowledge gap the subjects followed one practical session (\SI{95}{\minute}) during which they worked on an Angular application.
This preliminary session contained exercises on processing \gls{ui} events using the native Web API, Angular, and Interacto.
We asked the subjects to answer the three questions about their expertise after this session.
We asked the subjects not to talk to or help each other during the session.
The subjects were free to use any other resource to perform the tasks (online documentation, \etc).

\bigskip

\noindent \textbf{Results.}
The TIME, CORR, and DIFF results do not follow a normal distribution.
For all the statistical tests used in this section we consider a \pc{95} confidence level (\ie $p$-value$<$0.05).
\smallskip

\begin{table}[h]\small
	\centering\setlength{\tabcolsep}{4.5pt}
	\caption{Total test results, effect size, and confidence of the correctness results}\label{tab.res.corr}
	\begin{tabular}{cccSc}
		\toprule
		\textbf{Task}& \textbf{Tests} & \textbf{Tests} & \textbf{Odds Ratio}& \textbf{\emph{p}-value}\\
		& \textbf{Interacto}& \textbf{Angular}&  &\\
		& \textbf{(pass/fail)}& \textbf{(pass/fail)}&  &\\
		\cmidrule(lr){1-1}\cmidrule(lr){2-2}\cmidrule(lr){3-3}\cmidrule(lr){4-4}\cmidrule(lr){5-5}
		1 & \num{64} / \num{16} & \num{55} / \num{27}  & 1.95& \num{0.076}\\
		\cmidrule(lr){1-1}\cmidrule(lr){2-2}\cmidrule(lr){3-3}\cmidrule(lr){4-4}\cmidrule(lr){5-5}
		\rowcolor{lightgray}
		2 & \num{32} / \num{12} & \num{22} / \num{22} &2.63& \num{0.048}\\
		\cmidrule(lr){1-1}\cmidrule(lr){2-2}\cmidrule(lr){3-3}\cmidrule(lr){4-4}\cmidrule(lr){5-5}
		3 & \num{19} / \num{35} & \num{35} / \num{39} &0.61& \num{0.206}\\
		\cmidrule(lr){1-1}\cmidrule(lr){2-2}\cmidrule(lr){3-3}\cmidrule(lr){4-4}\cmidrule(lr){5-5}
		Total & \num{115} / \num{63} & \num{112} / \num{88} &1.43& \num{0.093}\\
		\cmidrule(lr){1-1}\cmidrule(lr){2-2}\cmidrule(lr){3-3}\cmidrule(lr){4-4}\cmidrule(lr){5-5}
		\rowcolor{lightgray}
		T-UNDO & \num{31} / \num{15} & \num{26} / \num{42} &3.3& \num{0.004}\\
		\cmidrule(lr){1-1}\cmidrule(lr){2-2}\cmidrule(lr){3-3}\cmidrule(lr){4-4}\cmidrule(lr){5-5}
		T-UI & \num{84} / \num{48} & \num{86} / \num{46} &0.93& \num{0.898}\\
		\bottomrule
	\end{tabular}
\end{table}

\Cref{tab.res.corr} reports the results regarding the correctness.
The columns '\emph{Tests}' reported the total number of executed tests.
From the results of the tests executions we produced a $2 \times 2$ contingency table.
We thus used the Fisher Exact Test~\cite{Sheskin2007} to study whether the results from G1 and G2 are independent.
To measure the size effect, we used the Odds Ratio~\cite{Sheskin2007}.
We did not consider the test results related to T1.2 or T3.2 for the subjects that did not start working on each of these task.

The results of T1 tend to be in favor of \wb, but are not significant enough to draw conclusions.
Regarding T2, the results are in favor of \wb (significant results) with an odds ratio of \num{2.63}:
on T2, the odds that a student achieves better correctness are increased by \pc{163} using \wb compared to Angular.
Regarding T3, the results do not exhibit significant results.
For T3.1, most G2 subjects used the native DnD feature that permits to graphically move objects:
Using \wb, G1 subjects had to use an \wb DnD to do such a move programmatically, which is a more complex task.
Similarly to T1, the total correctness tends to be in favor of \wb, but is not significant enough to draw conclusions.

Regarding the transversal task \emph{T-UNDO} the results are in favor of \wb (significant results) with an large effect size (\num{3.3}):
on undo/redo features, the odds that a student achieves better correctness are increased by \pc{230} using \wb compared to Angular.
Regarding \emph{T-UI} the results are not significant enough to draw conclusions.
\smallskip

\textbf{RQ5.1 conclusion.}
First, \wb helped the subjects in correctly adding undo/redo features to the application.
Second, we think that an \emph{entrance barrier} may exist for students to master \wb usages.
This may concern in particular the use of complex user interactions such as the DnD in T3.

\smallskip

\begin{table}[h]\small
	\centering\setlength{\tabcolsep}{1.6pt}
	\caption{Means, effect size, and confidence of the time results}\label{tab.res.time}
\begin{tabular}{ccccSc}
\toprule
\textbf{Task}& \textbf{Mean} 	 & \textbf{Mean}   & \textbf{Means Diff.} 	  & \textbf{\emph{Â}12}& \textbf{\emph{p}-value}\\
& \textbf{Interacto}& \textbf{Angular}& \textbf{(\% (minutes))}  &  &\\
& \textbf{(minutes)}& \textbf{(minutes)}&  &  &\\
\cmidrule(lr){1-1}\cmidrule(lr){2-2}\cmidrule(lr){3-3}\cmidrule(lr){4-4}\cmidrule(lr){5-5}\cmidrule(lr){6-6}
1.1 & \num{16.96} & \num{20.42} & \pc{-16.9} (\num{-3.46})&0.59& \num{0.442}\\
\cmidrule(lr){1-1}\cmidrule(lr){2-2}\cmidrule(lr){3-3}\cmidrule(lr){4-4}\cmidrule(lr){5-5}\cmidrule(lr){6-6}
\rowcolor{lightgray}
1 & \num{27.27} & \num{31.97} & \pc{-14.7} (\num{-4.7})&0.70& \num{0.045}\\
\cmidrule(lr){1-1}\cmidrule(lr){2-2}\cmidrule(lr){3-3}\cmidrule(lr){4-4}\cmidrule(lr){5-5}\cmidrule(lr){6-6}
\rowcolor{lightgray}
2 & \num{13.41}  & \num{18.48} & \pc{-27.4} (\num{-5.07})&0.81& $< 0.001$ \\
\cmidrule(lr){1-1}\cmidrule(lr){2-2}\cmidrule(lr){3-3}\cmidrule(lr){4-4}\cmidrule(lr){5-5}\cmidrule(lr){6-6}
\rowcolor{lightgray}
3.1 & \num{38.64} & \num{19.3} & +\pc{100.2} (+\num{19.34})&0.18&  $< 0.001$ \\
\cmidrule(lr){1-1}\cmidrule(lr){2-2}\cmidrule(lr){3-3}\cmidrule(lr){4-4}\cmidrule(lr){5-5}\cmidrule(lr){6-6}
3 & \num{37.65} & \num{38.37} & \pc{-1.9} (\num{-0.72})&0.44 & \num{0.484}\\
\cmidrule(lr){1-1}\cmidrule(lr){2-2}\cmidrule(lr){3-3}\cmidrule(lr){4-4}\cmidrule(lr){5-5}\cmidrule(lr){6-6}
\rowcolor{lightgray}
Total & \num{77.6} & \num{88.62} & \pc{-12.4} (\num{-11.02})& 0.85 & $< 0.001$\\
\cmidrule(lr){1-1}\cmidrule(lr){2-2}\cmidrule(lr){3-3}\cmidrule(lr){4-4}\cmidrule(lr){5-5}\cmidrule(lr){6-6}
\rowcolor{lightgray}
T-UNDO & \num{26.33} & \num{48.1} & \pc{-82.6} (\num{-21.77})& 0.71 & \num{0.029}\\
\cmidrule(lr){1-1}\cmidrule(lr){2-2}\cmidrule(lr){3-3}\cmidrule(lr){4-4}\cmidrule(lr){5-5}\cmidrule(lr){6-6}
T-UI & \num{68.3} & \num{58.28} & +\pc{17.2} (+\num{10.02})& 0.40 & \num{0.097}\\
\bottomrule
\end{tabular}
\end{table}

\Cref{tab.res.time} reports the results for the TIME variable.
We applied the Mann-Whitney test~\cite{Sheskin2007} as this test makes no assumptions about the distributions of assessed variables.
We measure the effect size using the Vargha-Delaney~\emph{Â} (\emph{Â}12) measure~\cite{vargha2000,arcuri2014} 
following \emph{Â}12 measure nomenclature~\cite{vargha2000}:
negligible: $>0.5$ small: $>0.56$, medium: $>0.64$, large: $>0.71$.
With the TIME variable, the less subjects spent time the better it is in favor of the used approach.

We did not report and discuss the results of T1.2 and T3.2 as their values may be strongly biased by the time limit of T1 and T2:
One subject may start T1.2 (or T3.2) several minutes before the limit affecting the relevance of studying T1.2 (or T3.2) alone.
The results of T1.2 and T3.2 are however integrated in the results of T1 and T3.

When concatenating the TIME results of T1, T2, and T3, the probability that the use of \wb would take less time than the use of Angular is \pc{85} (significant result).
This result is in favor of \wb and has a large effect size.

For T1, the probability that the use of \wb would take less time than the use of Angular is \pc{70} (significant result).
This result is thus in favor of \wb with a medium effect size.
The reason may be that using \wb the use of interactions is more direct.
Also, the support of undo/redo is native in \wb so that subjects did not had to implement their own undo/redo facilities.

For T2, the probability that the use of \wb would take less time than the use of Angular is \pc{81} (significant result).
This result is thus in favor of \wb and has a large effect size.
The G2 subjects had to deal with a widespread keyboard interaction not supported by most of \gls{ui} toolkits (Angular included), which takes time to support manually.
Similarly with T3.1, the probability that the use of \wb would take less time than the use of Angular is \pc{18} (significant result).
This result, in favor of Angular with a large effect size, has the same origin:
as we discussed with CORR, G2 subjects used the native Angular DnD that moves objects graphically. 
G1 subjects had to use the \wb DnD to change data, which may take more time.

For T3, the data do not show significant results in favor of \wb or Angular.
If T3.1 required less time for G2 as discussed in the previous paragraph, T3.2 required more time to G2 for coding undo/redo facilities.

\textbf{RQ5.2 conclusion.}
First, globally the use of \wb has a large positive impact of the time spent to do the three tasks against Angular.
Second, the results show that providing developers with user interactions has a benefit (for \wb in T2 or Angular in T3.1) in terms development time.

\smallskip

\begin{table}[h]\small
	\centering\setlength{\tabcolsep}{2.5pt}
	\caption{Means, effect size, and confidence of the difficulty results}\label{tab.res.diff}
\begin{tabular}{ccccSc}
\toprule
\textbf{Task}& \textbf{Mean} 	 & \textbf{Mean}   & \textbf{Means Diff.} 	  & \textbf{\emph{Â}12}& \textbf{\emph{p}-value}\\
& \textbf{Interacto}& \textbf{Angular}& \textbf{(\%)}  &  &\\
\cmidrule(lr){1-1}\cmidrule(lr){2-2}\cmidrule(lr){3-3}\cmidrule(lr){4-4}\cmidrule(lr){5-5}\cmidrule(lr){6-6}
1.1 & \num{3.27} & \num{3.95} & \pc{-17.2} (\num{-0.68})&0.56& \num{0.51}\\
\cmidrule(lr){1-1}\cmidrule(lr){2-2}\cmidrule(lr){3-3}\cmidrule(lr){4-4}\cmidrule(lr){5-5}\cmidrule(lr){6-6}
1.2 & \num{4.66} & \num{4.85} & \pc{-3.9} (\num{-0.19})&0.56& \num{0.54}\\
\cmidrule(lr){1-1}\cmidrule(lr){2-2}\cmidrule(lr){3-3}\cmidrule(lr){4-4}\cmidrule(lr){5-5}\cmidrule(lr){6-6}
\rowcolor{lightgray}
2 & \num{4.08}  & \num{5.76} & \pc{-29.2} (\num{-1.68})&0.69& \num{0.03}\\
\cmidrule(lr){1-1}\cmidrule(lr){2-2}\cmidrule(lr){3-3}\cmidrule(lr){4-4}\cmidrule(lr){5-5}\cmidrule(lr){6-6}
\rowcolor{lightgray}
3.1 & \num{7.19} & \num{4.13} & +\pc{74.1} (+\num{3.06})&0.26& $< 0.01$\\
\cmidrule(lr){1-1}\cmidrule(lr){2-2}\cmidrule(lr){3-3}\cmidrule(lr){4-4}\cmidrule(lr){5-5}\cmidrule(lr){6-6}
3.2 & \num{6.16} & \num{7.53} & \pc{-18.2} (-\num{1.37})&0.71& \num{0.18}\\
\bottomrule
\end{tabular}
\end{table}

\Cref{tab.res.diff} reports the level of difficulty asked to each subject for each sub-task.
Similarly to RQ5.2 we applied the Mann-Whitney test~\cite{Sheskin2007}.
This table does not contain lines \emph{Total}, \emph{T-UNDO}, and \emph{T-UI}:
since multiple subjects did not started (and reported the level of) T1.2 or T3.2 we cannot sum the results of the three tasks, contrary with TIME.
With the DIFF variable, the less subjects found a task hard the better it is in favor of the used approach.
The significant results concern T2 and T3.1 and seem to be related to the use of predefined user interactions:
The probability that the subjects find the task easier with \wb is \pc{69} for T2 and \pc{26} for T3.1.

\medskip

Regarding qualitative data for this sub-RQ, we now discuss some of the written comments of the subjects.
We identified five categories of remarks, mainly from G1 (\wb).

\emph{Pros of Interacto}.
Multiple subjects detailed that they liked \wb for several reasons.
A subject from G2 wrote that "\emph{after we've been taught a bit of Interacto it seems obvious that it can help to do complex actions more easily than only with Angular}".
Other subjects wrote that:
"\emph{Interacto significantly simplifies certain tasks}";
"\emph{Interacto very easy to handle, very easy to manage the undo/redo}".

\emph{Cons of Interacto.}
"\emph{I do not like the fact I have to define a \emph{ViewChild} in the HTML, then a property in the Angular component and then make the bind. I prefer the Angular system where you just have to bind a function in the HTML}".
We agree with this remark and already discussed it with RQ3 in \Cref{eval.case}.

"\emph{I also find that for very simple things it can quickly make the application heavier because it makes you write a lot of code.}"
Indeed, the coding of undoable classes in particular may take time.
A concrete benefit may appear when a developer reuses the same undoable command for several user interactions, which is a common case during the development of \glspl{ui} (\eg key shortcuts, menus).

\emph{Entrance barrier.}
"\emph{I think Interacto is very powerful, when you know how to use it well}".
"\emph{Struggling to understand totally \emph{dndBinder()}}".
"\emph{Once we understand the principle of the different \emph{Binder}, the code is done without too much difficulty}".
"\emph{Knowing what the variable \emph{i} was in the routines was the most useful part that I saw just at the end}".
All these remarks pointed out that using \wb requires learning its major concepts, which can hardly be done in two practical sessions.

\emph{Documentation issue.}
"\emph{The presentation of the Interacto documentation is not very intuitive.}"
"\emph{The user community is not yet very developed and therefore no help is available on the forums or on the net in general}".
Related to the previous point, these remarks stressed out that \wb has no user community compared to Angular so that it highly relies on its official documentation.

\smallskip

\textbf{RQ5.3 conclusion.}
First, the level of difficulty reported by the subjects confirmed that the use of predefined user interactions simplifies the coding activity.
Second, the informal feedback are mainly positive regarding the benefits of \wb.
The subjects, however, noticed various issues, in particular the entrance barrier that may exist to use \wb correctly.

\smallskip

\textbf{RQ5 conclusion.
This experiment with students on three representative user interaction development tasks exhibits several points.
First, the use of \wb is beneficial, in terms of time and correctness, for students to add undo/redo features to the application.
Second, the use of predefined user interactions, from both Angular and \wb, is also beneficial in terms of time and correctness.
This give another path to explore to improve \wb: 
\wb may also provide developers with predefined yet partial \wbers to do standard actions, similar to the Angular's DnD.
Third, an \emph{entrance barrier} to correctly use \wb may exist.
This barrier may concern the understanding about how to use the API to turn the execution of an interaction into a command.
This is normal for a novel technique and does not hamper the possible adoption of \wb since software engineers tend to learn new frameworks regularly.
}

\subsection{Threats to validity}

\textbf{External validity.} 
This threat concerns the possibility to generalize our findings.
The real world use case we detailed in \Cref{eval.case} was developed by authors of this paper.
To mitigate this issue we conducted an experiment with \num{44} master students in computer science to discuss the pros and cons of \wb compared to the Web API/Angular.

Closely related, \SOFTWARE is developed in JavaFX.
To mitigate this issue, we discussed in \Cref{eval.scope} about other use cases we developed in Angular and JavaFX.
The experiment conducted in \Cref{eval.std} used \iang within an Angular application.

The benchmarks conducted in \Cref{eval.case} concerns JavaFX code.
Claiming that the Angular implementation has no overhead as well may require dedicated experiments.
However, we cannot identify any reason for having such a difference between these two platforms as they rely on similar concepts.
Similarly, the experiment of \Cref{eval.std} used an Angular application.
We selected Angular as it follows state-of-the-art practices in terms of front-end development.

We design the experiment using three tasks representative of what a UI developers can do to process user interactions.
The selected user interactions (keyboard interaction, simple mouse interaction, DnD) are widespread user interactions.
These three tasks are of different difficulties and cover different aspects of the processing of user interactions.

Regarding the population validity, all the subjects were volunteers.
We asked the subjects to
fill a questionnaire before the experiment on their knowledge in front-end development. 
We used those data to design two balanced groups of subjects.

\textbf{Construct validity.} 
This threat relates to the perceived overall validity of the experiments.
Regarding tiredness, the duration of the experiment was \SI{95}{\minute}.
We chose this duration as it is the standard duration of the practical sessions that these students follow.
The use of a time limit may introduce a threat in the time results analysis.
We consider that not considering a time limit would lead to a more problematic threat to validity: the tiredness as previously discussed.
We do not consider the use of a time limit with students as an issue since by their current situation students got accustomed to time limited exercises.

Regarding the learning gap, none of the subjects knew \wb.
We conducted an initial practical sessions to introduce Angular and \iang concepts (data binding, \gls{ui} event processing).
Despite this effort, we are aware that subjects may feel more comfortable with Angular than with \iang.
This issue is accentuated with the documentation:
we noted that G1 subjects (\wb) strongly relied on the \wb documentation as the only source of information.
The G2 subjects could relied on various sources of information (Angular website, \emph{Stack Overflow}, \etc).

\section{Related Work}\label{sec.related}

We grouped the related research work into three categories:
approaches related to reactive programming and complex event processing;
approaches related to \gls{ui} event processing;
approaches that reify user interactions as first-class concepts.

\subsection{Reactive programming and complex event processing}

\gls{rp} provides abstractions and mechanisms to use time-changing values in programs~\cite{bainomugisha2013}.
In a user interface development context, \gls{rp} is used for two purposes:
binding and transforming (user interface) data;
processing \gls{ui} events, where \gls{ui} events are processed as data streams.
Data binding is broadly used by recent graphical toolkits (\eg Angular\footnote{\url{https://angular.io/guide/template-syntax}}, WPF\footnote{\url{https://docs.microsoft.com/en-us/dotnet/framework/wpf/data/data-binding-wpf}}, and Android\footnote{\url{https://developer.android.com/topic/libraries/data-binding}}) to update data on other data changes following \gls{rp} mechanisms.
The use of \gls{rp} to process \gls{ui} events brings several benefits compared to the use of callback methods.
First, it may reduce the size of the code thanks to various stream operators.
Second, it overcomes the identified limits of the \emph{Observer} pattern~\cite{maier2012deprecating,salvaneschi2014towards,Salvaneschi2014b,foust2015}.
We could use \gls{rp} to code the user interactions that Interacto provides, for example using the \emph{ReactiveX} library~\cite{meijer2010reactive,Maglie2016} that Angular already provides and that also works within JavaFX.
However, using \gls{rp} alone (\ie not within Interacto) to process \gls{ui} events has the following drawbacks that \wb overcomes as discussed in this paper:
\gls{ui} event is still the core concept and developers have to assemble events to build user interactions;
developers still have to write glue code to transform \gls{ui} events into commands.
The rest of this sub-section discusses the main \gls{rp} approaches related to user interface development.

ReactiveUI\footnote{\url{https://reactiveui.net}} is a \gls{rp} framework dedicated to the user interface development.
This framework considers commands as first-class concepts.
Developers can process \gls{ui} events by producing commands thanks to specific routines.
ReactiveUI, however, still consider \gls{ui} events as first-class concepts.

\cite{courtney2001} propose the use of \gls{rp} to develop user interfaces.
In particular, \gls{ui} events are considered as data streams that can be processed.
\gls{ui} events, however, are still first-class concepts in this approach.

\emph{Ur/Web} is a programming language for the Web~\cite{Chlipala:2015}.
\gls{rp} is used in this approach for rendering graphical objects.
Event callbacks are still used to control interactive objects.

\cite{Milicevic2013} propose a programming paradigm for developing interactive event-driven systems.
This approach is not specific to UIs.
When applied to user interfaces, it focuses on the rendering of graphical objects and \gls{ui} data binding.

\emph{Mobl} is a declarative language for programming mobile web applications~\cite{Hemel:2011}.
\emph{Mobl} implements the \emph{Model-View} pattern:
no controller is used to link views to data models.
The processing of low-level \gls{ui} events and the data bindings are moved to the \emph{View}.
One goal of \emph{Mobl} is to reduce the boilerplate code written in controllers to synchronize data models and views.
\emph{Mobl} promotes separation of concerns by supporting the separation of user interface and data model, which is a corner-stone of user interface engineering.
\emph{Mobl} provides reactive behavior mechanisms to be used directly in views to update them.
\wb is not tied to a specific architecture:
it requires accesses to the interactive objects that compose the user interface to process their events.
The current implementations of \wb, however, cannot work with the \emph{Model-View} pattern as views are usually described in an XML dialect.

\emph{Elm} is a functional \gls{rp} framework for programming user interfaces~\cite{czaplicki2013}.
When used to develop user interfaces, \emph{Elm} focuses on two aspects:
building and layouting user interfaces;
processing \gls{ui} events.
Similarly to the other \gls{rp} approaches, \emph{Elm} considers \gls{ui} low-level events only.
\emph{Elm} does not consider the concept of command.

\emph{Scala.React} is a Scala \gls{rp} framework that aims at overcoming the limits of the \emph{Observer} pattern~\cite{maier2012deprecating}.
This paper takes as example the development of user interfaces.
\gls{ui} events, however, are still first-class concepts and commands are not considered.

\emph{Flapjax} is a programming language for Web applications~\cite{meyerovich2009}.
\emph{Flapjax} provides, on the top of JavaScript, reactive programming features to tackle various web development problems.
As most of the languages or frameworks discussed in this section \emph{Flapjax} can be used to develop user interactions.
For example, the authors illustrate parts of \emph{Flapjax} by developing a DnD.
The authors notice that building user interactions bring benefits for developers:
"\emph{by separating the DnD event stream from action of moving the element, we have enabled a variety of actions}".
Commands, undo/redo, and the associated glue code, however, are not considered.

\cite{foust2015} discuss some challenges of programming user interfaces, in particular the data synchronization and update.
The authors highlight the complexity of understanding the spaghetti code provoked by the use of handlers to update data.
The authors propose an approach to overcome this problem.
This work focus in data binding and did not target \gls{ui} event processing.

Complex Event Processing (CEP) tackles the problem of analyzing data to detect event pattern~\cite{luckham2002}.
User interactions could be developed using CEP.
For example, a DnD may match the event-based pattern \emph{press-drag+-release}.
The main drawback is that a pattern is matched when all the required events are processed.
User interactions may require to process its events all along its execution.

\subsection{UI event processing approaches}

The Garnet system~\cite{Myers90} provides developers with a set of predefined, reusable, and customizable sets of behavioral interactive objects called interactors.
Interactors aim at hiding the \gls{ui} event processing from developers.
Following this work, the author of Garnet then proposed the use of (undoable) commands with interactors~\cite{Myers1996}.
These work are certainly the closest ones that inspired \wb.
They, however, follow a different philosophy than \wb.
Interactors are interactive objects predefined (yet customizable) for specific actions.
For example, the predefined \emph{move-grow} interactor aims at moving or changing the size of an object.
\wb promotes the concept of user interaction as a first-class concern to replace the current usage of \gls{ui} events.
By using \wb, developers are free to use a given user interaction to produce various commands.

More recently, \emph{InterState} is an approach for defining user interface behavior~\cite{oney2012constraintjs, oney2014interstate}.
The authors motivate the limits of the current \gls{ui} event-callback model to then propose the use of \glspl{fsm} to describe different parts of a user interface behavior.
Concretely, \emph{InterState} is a new programming language and environment for helping developers in coding and reusing \ui code.
With \wb we do not want developers to use another language or environment.
We aim at proposing developers with a technique that seamlessly works within their UI toolkits.
\emph{InterState} does not consider \gls{ui} commands.
Moreover, with \emph{InterState}, the states of a user interaction are directly bound to properties of a data object to change.
This makes the job of developers more complex when they have to replace the current user interaction with another one.

Based on the work of \cite{Myers90}, \cite{beaudoux:hal-00684881} propose an approach for developing interactive graphical objects.
This approach proposes a DSL (\emph{Domain-Specific Language}) embedded in Scala to develop interactors.
This work shares several ideas with the concept of \wb:
low-level \gls{ui} events are hidden from developers, to use predefined and customizable interactors, similarly to \cite{Myers90}.
These interactors can be controlled with some routines close to the ones that form the \wb fluent API, such as \emph{when}.
This approach, however, focuses on defining view templates.
It also does not propose a process to automatically transform user interactions into commands since commands are not first-class concepts in this approach.

\subsection{User interactions as objects}

Reifying user interactions as first-class concerns is not new and largely admitted in the \gls{hci} community.
\cite{appert2008} propose to code user interactions as \glspl{fsm} instead of using low-level \gls{ui} events.
This approach, however, fully focuses on user interactions and do not consider separation of concerns, code reuse, and \gls{ui} commands.

\cite{NAV09} propose the use of Petri nets to model the behavior of user interfaces, user interactions included.
This approach, however, does not propose any user interaction processing model.

\cite{blouin:inria-00477627,blouin:inria-00590891} propose an architectural design pattern where user interactions are modeled as \glspl{fsm}.
This design pattern also suggests the separation of concerns between user interactions and \gls{ui} commands.
No detail, however, is provided on a user interaction processing mechanism to transform such user interactions into commands.

Various UI modeling approaches aim at focusing on software interactivity.
Task modeling approaches, such as \emph{ConcurTaskTrees} (CTT)~\cite{paterno1997}, aim at expressing user's activities by describing tasks that users can do.
These approaches thus focus at a high level of abstraction on UI commands.
The \emph{Interaction Flow Modeling Language} (IFML)~\cite{brambilla2014} aims at specifying UIs.
IFML has an \emph{events specification} and an \emph{event transition specification} perspectives that detail the events (UI events included) that change the state of the UI and their impacts on this last.
UsiXML~\cite{limbourg2004} is a multi-level approach for building multi-platform and adaptive UIs.
The top level of UsiXML concerns task modeling with the same goals than CTT.
Finally, improving the interactivity or more generally the user experience of model-driven approaches and environments is a major concern~\cite{blouin:hal-01120558,abrahao2017}.
For example, \cite{sousa2019} propose to complete generative DSL environment approaches with models dedicated to UIs and user interactions for improving the user experience.

\subsection{UI Toolkits}

Several UI toolkits provide developers with features for coding (undoable) UI commands.
For example, Java Swing already proposed to associate UI commands to simple widgets (\eg buttons) and provides a linear undoable history.
These features are now part of JavaFX.
WPF has a similar feature for binding commands to simple widgets.
To the best of our knowledge, Angular and Android rely on event handlers for processing UI events and do not have command features such as the ones provided by WPF.
\wb is inspired by these features and brings facilities (algorithms, a dedicated API, run-time optimizations, object-oriented properties, \etc) that help developers in coding how executions of a user interaction can produce command instances. 
Interacto improves: 1/ the expressiveness of the use of UI commands with simple widgets with the above-mentioned facilities and with multiple binder routines (when, log, consume, \emph{etc.}) to customize the command creation; 
2/ the support of more complex user interactions in their usages for producing commands.

Several UI toolkits have reuse facilities: 
Angular has \emph{Directives}\footnote{\url{https://angular.io/api/core/Directive}}, WPF has \emph{Behaviors}\footnote{\url{https://github.com/Microsoft/XamlBehaviorsWpf/wiki}}, and Android has \emph{Slices}\footnote{\url{https://developer.android.com/guide/slices}}.
These features are interesting as they permit developers to extend the behavior of widgets with new features and new properties.
However even when using these features, a developer still has to handle low-level UI events to produce commands and gets no specific support for turning user interaction executions into commands as in \wb.
These UI toolkits features can however be very useful for \emph{implementing} \wb.
For example, our Angular implementation of \wb partly relies on Angular directives as a reuse mechanism.
We believe that a port of \wb to WPF or Android might similarly rely on Behavior and Slice objects.

Regarding the testability of commands and of the production of command instances, core testing and UI testing frameworks (e.g. TestFX for JavaFX, Expresso for Android, or Mocha/Jasmine/Protractor for Angular) provide core features that enable us to write more complex test assertions and test frameworks: 
based on and complementary to these frameworks, \wb proposes dedicated testing oracles implemented in tools for easing the writing of tests for commands and production of commands.
Our implemented testing tool both generate skeletons of test classes and also provides a framework for helping in writing UI command tests and for testing the production of commands.

Finally, several UI toolkits try to overcome the problem of relying on basic \gls{ui} events only by providing supplementary \gls{ui} events.
For example, the HTML API provides the \emph{dragstart}, \emph{drag}, and \emph{dragend} events that respectively represent the starting, the running, and the ending of a \gls{dnd}.
If such \gls{ui} are a progress towards the support of complex user interactions, they still rely on the UI event processing model and its limits:
they do not help developer in turning UI events into commands, while \wb notably provides algorithms, a dedicated API, run-time optimizations, object-oriented properties, for this purpose.

\section{Conclusion}\label{sec.conclusion}

This paper presents \wb, a novel user interaction processing model.
Based on software engineering good practices, \wb aims at better engineering \uis.
Instead of providing developers with low-level \gls{ui} event processing, \wb reifies user interactions and \gls{ui} commands as first-class concerns.
The two implementations of \wb, \ijfx and \iang, show that the proposal is not tied to a specific language or \gls{ui} platform.

The long term experiment shows that the proposal scales for one very interactive and widely-used software system.
The experiment conducted with students exhibited several pros and cons of the proposal.
The use of \wb is beneficial, in terms of time and correctness, for students to add undo/redo features to the application.
The use of predefined user interactions is also beneficial in terms of time and correctness.
However, an \emph{entrance barrier} to use correctly \wb may exist.

In our future work, we will investigate how to provide developers with predefined yet partial \wbers to do standard actions, similar to the Angular's DnD that moves objects graphically.
Second, we will investigate how to help HCI designers in designing and testing novel user interactions, and how to produce concrete user interactions for integration in \wb.
Finally, \wb may also help the design of dynamic and static code analyzing techniques for the producing of UI tests, in the continuation of the test generation technique we propose.

\bibliographystyle{IEEEtran}
\bibliography{main}

\end{document}